\documentclass[aps,floatfix,twocolumn,prd,superscriptaddress,nofootinbib,floats,preprintnumbers]{revtex4-1}

\usepackage{amsmath}
\usepackage{amssymb}
\usepackage{mathrsfs}
\usepackage{graphicx}
\usepackage{hyperref}
\usepackage[T1]{fontenc}
\usepackage{slashed}
\usepackage{bbold}
\usepackage{color}
\usepackage{url}
\usepackage{verbatim}
\usepackage{tikz}
\usepackage{setspace}
\usepackage{placeins}
\usepackage{physics}
\usepackage{float}
\usepackage[normalem]{ulem}
\usetikzlibrary{arrows.meta,shapes.geometric,calc,shadings}
\tikzset{
  mynode/.style={
    execute at begin node=\setlength{\baselineskip}{1em}
  }
}

\allowdisplaybreaks
\newcommand{\beq}{\begin{eqnarray}}
\newcommand{\eeq}{\end{eqnarray}}
\newcommand{\bea}{\begin{eqnarray}}
\newcommand{\eea}{\end{eqnarray}}

\newcommand{\nnmb}{\nonumber}
\newcommand{\del}{\partial}
\newcommand{\lrf}[2]{\left(\frac{#1}{#2}\right)}

\newcommand{\kev}{\,{\text{keV}}}
\newcommand{\mev}{\,{\text{MeV}}}
\newcommand{\gev}{\,{\text{GeV}}}

\begin{document}

\preprint{MIT--CTP/5405}

\title{Accelerating Earth-bound dark matter}
\author{David McKeen}
\email{mckeen@triumf.ca}
\affiliation{TRIUMF, 4004 Wesbrook Mall, Vancouver, BC V6T 2A3, Canada}

\author{Marianne Moore}
\email{mamoore@mit.edu}
\affiliation{TRIUMF, 4004 Wesbrook Mall, Vancouver, BC V6T 2A3, Canada}
\affiliation{Department of Physics and Astronomy, University of British Columbia, Vancouver, BC V6T 1Z1, Canada}
\affiliation{Center for Theoretical Physics, Massachusetts Institute of Technology, Cambridge, Massachusetts 02139, USA}

\author{David~E.~Morrissey}
\email{dmorri@triumf.ca}
\affiliation{TRIUMF, 4004 Wesbrook Mall, Vancouver, BC V6T 2A3, Canada}

\author{Maxim Pospelov}
\email{pospelov@umn.edu}
\affiliation{School of Physics and Astronomy, University of Minnesota, Minneapolis, Minnesota 55455, USA}
\affiliation{William I. Fine Theoretical Physics Institute, School of Physics and Astronomy, University of Minnesota, Minneapolis, Minnesota 55455, USA}

\author{Harikrishnan Ramani}
\email{hramani@stanford.edu}
\affiliation{Stanford Institute for Theoretical Physics, Stanford University, Stanford, California 94305, USA}
\date{\today}
\begin{abstract}
A fraction of the dark matter may consist of a particle species that interacts much more strongly with the Standard Model than a typical weakly interacting massive particle~(WIMP) of similar mass. Such a strongly interacting dark matter component could have avoided detection in searches for WIMP-like dark matter through its interactions with the material in the atmosphere and the Earth that slow it down significantly before reaching detectors underground. These same interactions can also enhance the density of a strongly interacting dark matter species near the Earth's surface to well above the local galactic dark matter density. In this work, we propose two new methods of detecting strongly interacting dark matter based on accelerating the enhanced population expected in the Earth through scattering. The first approach is to use underground nuclear accelerator beams to upscatter the ambient dark matter population into a WIMP-style detector located downstream. In the second technique, dark matter is upscattered with an intense thermal source and detected with a low-threshold dark matter detector. We also discuss potential candidates for strongly interacting dark matter, and we show that the scenario can be naturally realized with a hidden fermion coupled to a sub-GeV dark photon.
\end{abstract}

\maketitle

\section{Introduction}

Dark matter~(DM) makes up 25\% of the energy budget of the Universe and manifests itself on many distance scales, from the cosmological horizon down to individual satellite galaxies~\cite{Planck:2018vyg,Navarro:1995iw}. Despite a plethora of evidence for DM based on its gravitational influence on visible matter, the identity of DM remains a mystery. The simplicity and success of the lambda-cold dark matter~($\Lambda$CDM) model of the early Universe~\cite{Planck:2018vyg}, where DM evidence is seen prior to the CMB decoupling, motivates ``elementary'' DM candidates. Many particle physics candidates have been suggested as prospective particle DM~\cite{Jungman:1995df,Bertone:2004pz,Feng:2010gw}, and a diverse set of experimental programs have been devised to detect them in a laboratory setting~\cite{Battaglieri:2017aum}. 

Approaches to DM include ``bottom-up'' phenomenological models of DM as well ``top-down'' scenarios where DM is part of a more complete framework that addresses other shortcomings of the Standard Model~(SM)~\cite{Jungman:1995df,Bertone:2004pz,Feng:2010gw}. A common element in nearly all these theories is nongravitational interactions between DM and Standard Model particles. Such interactions can mediate the production of DM in the early Universe from collisions within the SM plasma, providing an explanation for the large DM abundance observed today. These interactions also form the basis of most attempts to detect DM in the laboratory. 

An extremely robust experimental effort has been developed to detect DM in the form of weakly interacting massive particles~(WIMPs), where DM connects to the SM through the weak force. The standard detection technique is to search for WIMPs scattering off atomic nuclei or electrons in the target and imparting some of their kinetic energy to the detector. Large-scale detectors with energy thresholds down to $E_{\rm thr} \gtrsim \kev$ have put very strong bounds on weak-scale DM and significantly constrain or exclude many WIMP models of DM~\cite{Aprile:2018dbl,LUX:2016ggv,PandaX-4T:2021bab}. New detector developments have achieved even lower detection energy thresholds down to $E_{\rm thr} \sim \mathrm{eV}$, albeit with smaller detector masses and volumes~\cite{SENSEI:2020dpa,SuperCDMS:2020ymb,CRESST:2019jnq}.

While searches for WIMPs have strong motivation, it is also important to consider how to find other DM candidates that would not show up in standard direct detection experiments. In this paper, we investigate scenarios where some or all of the DM interacts ``strongly'' with visible matter, corresponding to the scattering length of DM in the Earth being small compared to the overburden of a typical direct detection experiment~\cite{Dimopoulos:1989hk,Starkman:1990nj,Farrar:2003gh,Zaharijas:2004jv}.\footnote{We use ``strongly interacting DM'' here to refer to DM that interacts with the SM with a large cross section, whether by quantum chromodynamics~\cite{Jaffe:1976yi,Farrar:2003gh,Zaharijas:2004jv,DeLuca:2018mzn} or through a new force carrier~\cite{Boehm:2003hm,Pospelov:2007mp}.} By exchanging momentum with particles in the Earth, strongly interacting DM particles quickly thermalize with the surrounding matter with kinetic energies $E\lesssim 0.05~\rm eV$, much lower than the detection threshold of nearly all underground WIMP detectors~\cite{Baxter:2021pqo}. Consequently, strongly interacting DM does not deposit sufficient energy to be detected by scattering in standard direct detection experiments~\cite{Starkman:1990nj,Collar:1993ss,Foot:2014osa,Davis:2017noy,Kavanagh:2017cru,Emken:2018run,Hooper:2018bfw,Emken:2019tni}. 

\begin{figure*}[ttt]
\centering
\begin{minipage}[b]{0.5\textwidth}
        \centering
        \includegraphics[height=5.705cm]{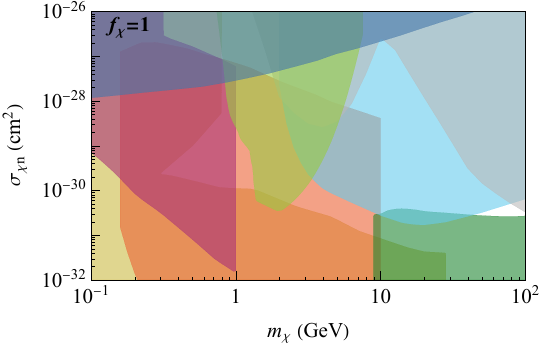} 
  \end{minipage}~~
  \begin{minipage}[b]{0.5\textwidth}
    \centering
    \includegraphics[height=5.705cm]{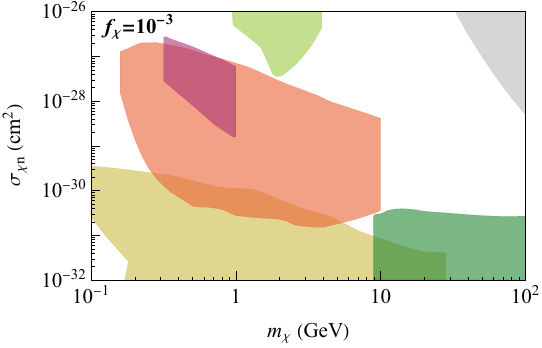}
  \end{minipage} 
\begin{minipage}[b]{0.5\textwidth}
        \centering
        \includegraphics[height=5.705cm]{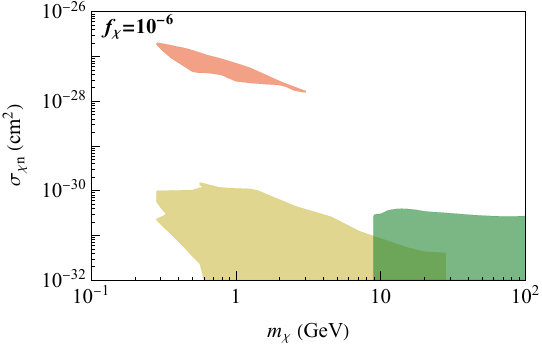} 
  \end{minipage}~~
  \begin{minipage}[b]{0.5\textwidth}
    \centering
    \includegraphics[height=5.705cm]{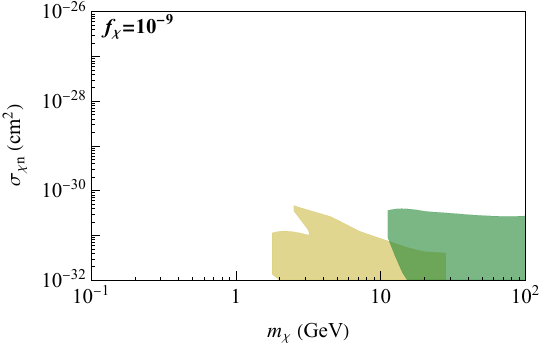}
  \end{minipage} 
  \includegraphics[width=.73\textwidth]{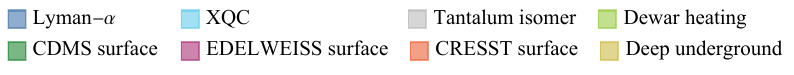}
\caption{Current bounds on a strongly interacting dark matter component $\chi$ that scatters with nuclei through a spin-independent interaction and that makes up a fraction $f_\chi = 1,10^{-3}, 10^{-6}, 10^{-9}$ of the total dark matter density as a function of its mass and effective per-nucleon cross section. Details of how these bounds are obtained and a list of the original experimental sources are given in Appendix~\ref{sec:appb}.
}
\label{fig:SIDMbounds}
\end{figure*}

Various strategies have been employed to test such strongly interacting DM. Indirect bounds have been derived by considering the impact of strongly interacting DM on primordial nucleosynthesis~\cite{Cyburt:2002uw}, the cosmic microwave background~\cite{Maamari:2020aqz,Buen-Abad:2021mvc}, Milky Way satellite galaxies~\cite{Maamari:2020aqz,Buen-Abad:2021mvc}, and large scale structure~\cite{Rogers:2021byl}. More direct limits have been obtained from DM searches at the surface of the Earth such as the CRESST~\cite{CRESST:2017ues} and EDELWEISS~\cite{EDELWEISS:2019vjv} surface runs as well as reanalyses~\cite{Wandelt:2000ad,Erickcek:2007jv,Mahdawi:2017cxz,Mahdawi:2018euy} of the rocket-based X-Ray Quantum Calorimeter~(XQC) detector~\cite{McCammon:2002gb}. In this case the column depth of material that the DM passes through on its way to the detector is far smaller so that it can impart considerable energy and be detected. However, the mitigation of the cosmic-ray-created backgrounds proves to be a considerable (but not insurmountable~\cite{Collar:2018ydf}) challenge. Lighter strongly interacting DM with mass $m_\chi \lesssim \gev$ can also be upscattered by cosmic rays to produce a detectable signal in near-surface neutrino detectors~\cite{Bringmann:2018cvk,Cappiello:2019qsw,Ema:2018bih} and bounds have been determined by PROSPECT~\cite{PROSPECT:2021awi} and PandaX-II~\cite{PandaX-II:2021kai}. Limits on the terrestrial population of such DM have been obtained from the heating it can cause in cryogenic Dewars~\cite{Neufeld:2018slx,Neufeld:2019xes,Xu:2021lmg} as well as the deexcitation of certain long-lived isomers of tantalum~(Ta) it can induce by scattering~\cite{Pospelov:2019vuf,Lehnert:2019tuw}. In Fig.~\ref{fig:SIDMbounds} we summarize existing bounds on a strongly interacting DM candidate $\chi$ that makes up a fraction $f_\chi = \rho_{\chi}/\rho_\text{DM}$ of the total cosmological DM density and that scatters with nuclei through a spin-independent (SI) interaction. More details on these exclusions and their experimental basis can be found in Appendix~\ref{sec:appb}.

In this paper, we propose an alternative strategy for detecting strongly interacting DM based on upscattering that leverages precisely the same tendency of this type of DM to thermalize that makes it so challenging to detect deep underground. Specifically, the robust interactions between DM and the Earth lead to efficient capture, inhibited evaporation, and slow propagation through the crust. Together, this generates an enhanced abundance of strongly interacting DM within the Earth, both in terms of an equilibrated thermal population~\cite{Neufeld:2018slx} as well as a transient sinking population~\cite{Pospelov:2019vuf,Pospelov:2020ktu}. These large DM densities open the possibility of an active ``two-step'' detection process where SM projectiles are used to \emph{upscatter} DM within the Earth to energies that are observable in nearby detectors. 

We consider two realizations of this approach in the present work. First, we investigate DM searches using existing or planned nuclear accelerators installed in deep underground labs for precision studies of nuclear reactions in low radiation environments. These include the LUNA accelerator currently deployed at the Laboratori Nazionale di Gran Sasso~(LNGS) featuring proton and helium ion beams with current $I_b= 1\,\text{mA}$ and terminal accelerating beam voltage $V_b = 0.4\,\text{MV}$~\cite{FORMICOLA2003609,Costantini:2009wn}. Upgrade plans include LUNA-MV at LNGS with $V_b=0.5$--$3.5\,\text{MV}$, $I_b = 0.1$--$1.0\,\text{mA}$ and beams of protons, helium ions, and carbon ions~\cite{SEN2019390,Prati:2020uxj}, JUNA at the China JinPing underground Laboratory~(CJPL) with $V_b = 0.4\,\text{MV}$, $I_b = 2.5$--$12\,\text{mA}$ and beams of protons and helium ions~\cite{juna}, CASPAR at the Sanford Underground Research Facility~(SURF) with $V_b = 1.1\,\text{MV}$, $I_b=0.25\,\text{mA}$ and beams of protons and helium ions~\cite{caspar}, and possibly even higher power, accelerators with gradients above $V_b \gtrsim 10\,\text{MV}$ dedicated to neutrino physics~\cite{Abs:2015tbh}. These ion beams can upscatter DM to energies in the $\kev$ range that standard WIMP DM detectors are most sensitive to; therefore a detector placed downstream could potentially detect this DM. Given the fact that the same labs usually host a variety of DM experiments, an implementation of such a detection scheme appears to be feasible. 

The second approach that we study is the upscattering of DM by gas atoms in a hot thermal source. DM that is thermalized near the Earth's surface has kinetic energies near $0.025~{\rm eV}$, well below the threshold of any near-future direct detection experiment. However, in the presence of a thermal source containing gas with temperature $T \sim 3000\,\text{K}$,\footnote{Recent examples of this technology include incandescent light bulbs.} DM near the Earth's surface can be upscattered to kinetic energies around a $\rm eV$, potentially within reach of planned detectors. While this presents a challenging signal, it is not outside the realm of possibility, with existing experiments already probing electron recoils with $\Delta E \sim \text{eV}$~\cite{SENSEI:2020dpa,SuperCDMS:2020ymb,CRESST:2019jnq} and future proposals extending down to energy depositions below $\Delta E \sim 10\,\text{meV}$~\cite{Hochberg:2015pha,Hochberg:2015fth,Schutz:2016tid,Hochberg:2017wce,Knapen:2017ekk}. This technique could also be further improved by optimizing the thermal sources.

Following this introduction, we discuss the capture, slowing, and resulting density distribution of strongly interacting DM in the Earth in Sec.~\ref{sec:capture}. Next, in Sec.~\ref{sec:acc} we compute the upscattering and detector event rates of strongly interacting DM in deep underground accelerator facilities for a hypothetical detector placed downstream of the beam. In Sec.~\ref{sec:hot}, we present estimates for event rates in low-threshold detectors from DM upscattered by a thermal source. In Sec.~\ref{sec:models}, we discuss specific realizations of strongly interacting DM. Finally, Sec.~\ref{sec:conc} is reserved for our conclusions. A review of bounds on strongly interacting DM is given in Appendix~\ref{sec:appb}, while technical details about thermal upscattering are presented in Appendix~\ref{sec:appa}.

\section{Overview of Dark Matter Accumulation\label{sec:capture}}

As the Earth passes through the local DM halo it can interact with and trap DM particles~\cite{Press:1985ug,Gould:1987ju,Gould:1987ir,Gould:1988eq}. This occurs when DM scatters with material in the Earth such that its velocity falls below the local escape velocity. Once captured, a strongly interacting DM particle will undergo further scatterings to thermalize and sink until it reaches an equilibrium distribution within the Earth. Over the history of the Earth, this leads to a thermal (or \emph{Jeans}) population of DM within it. In addition to the thermal population, an itinerant \emph{traffic jam} population can arise when the sinking rate of DM is slow relative to its capture rate~\cite{Pospelov:2019vuf,Lehnert:2019tuw,Pospelov:2020ktu}. We present our estimates for both populations in this section.

\subsection{Thermal density}

The thermal density is the population of DM captured by the Earth over its history that has thermalized~\cite{Gould:1987ir,Gould:1988eq}. When DM interacts strongly with the Earth, several new qualitative features arise relative to more weakly interacting scenarios, as emphasized in Ref.~\cite{Neufeld:2018slx}. In particular, DM is expected to reach local thermal equilibrium with the baryonic matter around it, and thermal evaporation of DM from the Earth is impeded by scattering. Together, these features can lead to a significant thermal population of captured DM near the surface of the Earth~\cite{Neufeld:2018slx}.

For a DM particle $\chi$ that does not annihilate significantly, the total number of $\chi$ particles captured in the Earth over its history is~\cite{Gould:1987ju}
\beq
    N_\chi = \frac{C}{E}\,\left(1 - e^{-E\,t_\oplus}\right) \ ,
\eeq
where $t_\oplus \simeq 4.54\times 10^9\,\text{yr}$ is the age of the Earth, $C$ is the total capture rate, and $E$ is the loss rate due to evaporation. We apply the calculations of Ref.~\cite{Neufeld:2018slx} specific to strongly interacting DM to estimate the rates of capture and evaporation. 

The capture rate for the large cross sections that we study here is well approximated by the Earth geometric cross section times the local DM flux up to small corrections~\cite{Neufeld:2018slx},
\beq
C = f_{\rm cap}\,\pi\,R_{\oplus}^2\,\frac{\rho_\chi}{m_\chi}\,v_{\rm eff} \ .
\label{eq:cap}
\eeq
Here, $R_\oplus \simeq 6370\,\text{km}$ is the Earth radius, $v_{\rm eff} \simeq 250\,\text{km/s}$ is the effective local DM flux velocity, and $f_{\rm cap}$ is the fraction of DM that is captured when it strikes the Earth. The capture fraction is less than unity because some of the DM that encounters the Earth will be reflected and is estimated to be~\cite{Neufeld:2018slx}
\beq
f_{\rm cap} = \left[\frac{4}{\pi}
\frac{\ln(1-\beta_\chi)}{\ln(v_{\rm eff}^2/v_{\rm esc}^2)}\right]^{1/2} \ ,
\eeq
where $v_{\rm esc}(R_\oplus) \simeq 11.2\,\text{km/s}$ is the surface escape velocity and $\beta_{\chi} = 2m_\chi m_N/(m_\chi+m_N)^2$ is the mean fractional energy loss per collision with nucleus $N$. Up to the factor of $f_{\rm cap}$, this capture rate matches well with other approaches~\cite{Bernal:2012qh,Garani:2021feo}.

Evaporation of DM arises from thermal collisions that upscatter DM particles beyond the local Earth escape velocity~\cite{Gould:1987ju,Gould:1987ir}. When the interaction length of DM is much smaller than the radius of the Earth, this process is strongly impeded by the rescattering of DM before it exits the Earth or the atmosphere. Assuming a DM number density profile of $n_\chi(r)$ and local velocity $v_{\text{th},\chi}(r)$ in the Earth, the evaporation rate from Ref.~\cite{Neufeld:2018slx} is 
\beq
E = \left.\frac{n_\chi\, v_{\text{th},\chi}}{2\,\sqrt{\pi}}
\left(1+\frac{v_{\rm esc}^2}{v_{\text{th},\chi}^2}\right)
e^{-v_{esc}^2/v_{\text{th},\chi}^2}\right|_{r_{\rm LSS}} \ ,
\eeq
where all quantities are to be evaluated at the last-scattering radius $r_{\rm LSS}$. This is obtained from the condition
\beq
1 = \int_{r_{\rm LSS}}^\infty\!dr\;\sum_N\sigma_{\chi N}\,n_N(r) \ ,
\eeq
with $n_N(r)$ the local number density of species $N$.
 
Determining the DM evaporation rate requires the radial distributions of its mean velocity $v_{\text{th},\chi}(r)$ and its number density $n_{\chi}(r)$ within the Earth and the atmosphere. The mean velocity is just the local thermal one, $v_{\text{th},\chi}(r) = \sqrt{8\,T(r)/\pi\,m_\chi}$ where $T(r)$ is the local temperature. For the number density, we follow Ref.~\cite{Neufeld:2018slx} and obtain it by balancing the local DM thermal pressure $p_\chi = n_\chi\,T$ with gravity,
\beq
\frac{1}{n_\chi\,T}\,\frac{d(n_\chi\,T)}{dr} = - \frac{G\,m_\chi\,M_r}{r^2\,T} \ ,
\eeq
where $T=T(r)$, and $M_r$ is the total mass enclosed at radius $r$ from the center of the Earth. Note that this expression is based on the assumption of local thermal equilibration, which is expected to occur for the range of large cross sections of interest. We integrate this expression using the PREM Earth density and temperature profiles from Refs.~\cite{Anzellini,prem} and the NRLMSISE-00 model of the atmosphere~\cite{NRLMSISE} and set $n_\chi(\infty)=n_\chi^\text{halo}$.

Our estimates for the thermal density of captured DM are subject to a number of uncertainties whose total impact we expect to be small in terms of the regions in the $m_\chi$-$\sigma_{\chi n}$ plane where the density is strongly enhanced. The most significant one is our simplified treatment of evaporation following Ref.~\cite{Neufeld:2018slx}, which neglects the possibility of DM escaping from inside the last scattering surface, $r < r_\text{LSS}$. This would tend to push the masses for which evaporation strongly depletes the thermal DM density to slightly higher values. However, we find that the depletion due to evaporation varies very rapidly with mass, and thus, the shift in mass will be mild. A further approximation is our assumption of thermal equilibration. We find that this is a good approximation when the DM density is significantly enhanced near the Earth's surface over the halo value since this requires large cross sections and, therefore, frequent scattering with material in the Earth, to avoid major losses to evaporation.

\subsection{Traffic jam density}

In addition to the thermal population of accumulated dark matter, large DM cross sections also give rise to an itinerant traffic jam population consisting of infalling DM that has not yet reached its equilibrium distribution~\cite{Pospelov:2019vuf,Pospelov:2020ktu}. The local density of this component can be obtained from the steady-state conservation of DM flux in the form
\beq
\frac{n_\chi (r)}{\rho_{\chi}/m_\chi} = \lrf{R_\oplus}{r}^2\frac{\langle v_z\rangle}{v_z(r)} \ ,
\eeq
where $\langle v_z\rangle \simeq v_{\rm eff}/2$ is the mean inward flux velocity of local DM beyond the Earth and $v_z(r)$ is the average downward DM velocity at radius $r$. Obtaining $n_\chi(r)$ is therefore just a matter of computing $v_z(r)$.

DM entering the Earth is pulled inward by gravity while being slowed by collisions with the surrounding material. To describe these effects, it is convenient to define the inward coordinate $z = R_\oplus - r$ and split the evolution of $v_z$ into
two regimes: $v > v_{\text{th},\chi}$ and $v < v_{\text{th},\chi}$, where $v \geq v_z$ is the DM speed and $v_{\text{th},\chi} \simeq \sqrt{8\,T/\pi\,m_\chi}$ is the local thermal DM speed. In both regimes, we find that the downward velocity is well described by the relation
\beq
v_z\,\frac{dv_z}{dz} = g - v_z\,\sum_N\lrf{\mu_N}{m_\chi}n_N\,\sigma_{T,N}\,\tilde{v}_N\ ,
\label{eq:vz}
\eeq
with the sum running over all relevant targets $N$ with mass $m_N$, $\mu_N = m_N m_\chi/(m_N+m_\chi)$ is the reduced mass, $n_N$ is the target number density, $\sigma_{T,N}$ is the transfer cross section, and $\tilde{v}_N = \max\{v,\,v_{\text{th},N}\}$ where $v_{\text{th},N} \simeq \sqrt{8\,T/\pi\,m_N}$ is the relevant thermal velocity of the target. Evaluating Eq.~\eqref{eq:vz} also requires that we specify the DM speed $v$. For $v > v_{\text{th},\chi}$, the speed is reduced by collisions and can be tracked through the rate of energy loss~\cite{landau1981course} such that
\beq
v_z\,\frac{dv}{dz} = -\sum_N\lrf{\mu_N^2}{m_Nm_\chi}n_N\,\sigma_{T,N}\,\frac{\tilde{v}_N^3}{v} \ .
\label{eq:vv}
\eeq
This expression neglects gravity, which is a good approximation for the parameters of interest in this work. Note as well that $v$ and $v_z$ have different kinematic dependencies on the masses for $m_\chi \ll m_N$. In this limit, multiple collisions are needed to reduce $v$ by a significant amount even though a single collision can reflect the DM particle and change $v_z$ by order unity. We apply Eq.~\eqref{eq:vv} to estimate the slowing of $v$ down to $v_{\text{th},\chi}$. Once $v \to v_{\text{th},\chi}$, further collisions are not expected to change its speed on average, and thus, we fix $v = v_{\text{th},\chi}$ from this point on.

As for the calculation of the enhanced thermal density, our results for the traffic jam density are subject to several uncertainties. The main sources here are our neglect of reflection of incoming DM and evaporation once it reaches a steady state. These only become significant at lower masses $m_\chi \lesssim 3\,\gev$, and for this reason we do no extend our traffic jam calculation below this value.

\subsection{Enhanced densities}

\begin{figure}[ttt]
    \centering
    \includegraphics[width=0.413\textwidth,trim=0.05cm 0 0 0,clip]{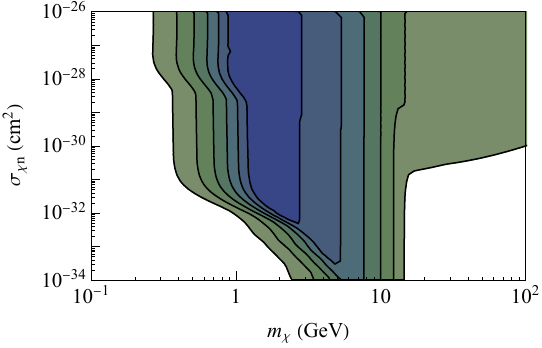}
    \raisebox{0.12\height}{\includegraphics[height=4.2cm,trim=0.235cm 0 0.24cm 0,clip]{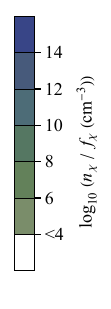}}
    \caption{Enhanced density of a strongly interacting dark matter species $\chi$ at a depth of $1.4$ km under the surface of the Earth as a function of mass~($m_\chi$) and per-nucleon cross section~($\sigma_{\chi n}$), with both thermal and traffic jam populations included. The dark matter-nuclear interaction is assumed to be spin independent with equal couplings to protons and neutrons.}
\label{fig:ndm}
\end{figure}

In Fig.~\ref{fig:ndm} we show the enhanced density of DM at a depth of $z = 1.4$ km under the Earth, corresponding to the overburden at LNGS, as a function of mass $m_\chi$ and per-nucleon cross section $\sigma_{\chi n}$. Both the thermal and traffic jam contributions to the enhanced density are included. To connect the per-nucleon cross section to cross sections on nuclei, we take $\sigma_{T,N} = \min\{A^2\,(\mu_{\chi N}/\mu_{\chi n})^2\sigma_{\chi n},\,4\,\pi\,r_N^2\}$, where $A$ is the atomic mass of nucleus $N$, and $r_N \simeq (1.2\,\text{fm})\,A^{1/3}$ is the nuclear radius. This form corresponds to a SI point interaction with a nuclear form factor of unity together with a saturation of the cross section at the geometric area of the nucleus~\cite{Pospelov:2019vuf,Digman:2019wdm}. Since most of the scatterings leading to capture and accumulations have a low momentum transfer relative to the inverse nuclear radii $1/r_N$, we expect that setting the nuclear form factor to unity should be a good approximation.

The DM densities shown in Fig.~\ref{fig:ndm} are much larger than the local halo density, particularly for larger cross sections. This enhancement has two primary features corresponding to the thermal and traffic jam components, respectively. The greatest enhancement between $m_\chi \sim 1$--$10~\gev$ comes from thermal accumulation and coincides with that found in Ref.~\cite{Neufeld:2018slx}. The density enhancement from this component at the locations of other underground accelerators is very similar. Evaporation depletes this population for $m_\chi \lesssim 1\,\gev$, while for $m_\chi \gtrsim 10\,\gev$, the thermal population is mainly located deeper within the Earth.\footnote{Since we do not consider evaporation effects in our traffic jam calculations, we only include this component of the enhanced density for $m_\chi \geq 3\,\gev$.} Instead, the dominant enhancement at larger masses $m_\chi \gtrsim 10\,\gev$ comes from the traffic jam population. If $\chi$ makes up only a fraction $f_\chi$ of the total DM energy density, the densities shown in Fig.~\ref{fig:ndm} are reduced by the same factor.

\section{Upscattering of dark matter by accelerator beams\label{sec:acc}}

In this section we investigate the upscattering of strongly interacting dark matter by the beams of deep underground accelerators such as LUNA~\cite{FORMICOLA2003609,Costantini:2009wn}, LUNA-MV~\cite{SEN2019390,Prati:2020uxj}, JUNA~\cite{juna}, and CASPAR~\cite{caspar}. We compute the upscattering rates as well as the detection rates through elastic nuclear scattering in a xenon detector of modest size.

\subsection{Dark Matter Upscattering by Accelerator Beams}

Consider a beam of nuclei of mass $m_b$ and kinetic energy $E_b \ll m_b$ incident on a cloud of DM particles $\chi$ effectively at rest. If a beam nucleus collides with a DM particle in the cloud, the DM will be upscattered to a velocity
\beq
v_\chi = \lrf{2\mu_{\chi b}}{m_\chi}\sqrt{\frac{2E_b}{m_b}}\,\cos\theta \ ,
\eeq
where $\theta$ is the angle of the outgoing DM relative to the beam direction. Should the upscattered DM particle collide with a target nucleus $N = (A,Z)$ in a detector, the nucleus will recoil with kinetic energy 
\beq
E_R &=& \frac{({2\,\mu_{\chi N}}\,v_\chi\,\cos\alpha)^2}{2\,m_N} \\
&=& E_b\lrf{2\mu_{\chi N}}{m_N}\!\lrf{2\mu_{\chi N}}{m_\chi}\!\lrf{2\mu_{\chi b}}{m_b}\!\lrf{2\mu_{\chi b}}{m_\chi}\cos^2\theta\;\cos^2\alpha
\nnmb\\
&\equiv& E_{R}^{\rm max}\,\cos^2\alpha \ ,
\nnmb
\eeq
where $\alpha$ is the angle between the recoiling nucleus and the incident DM direction. We note that all the factors multiplying $E_b$ in this expression are less than unity and represent the combined kinematic suppression from the two scattering reactions involved. 

\begin{figure*}[ttt]
\centering

\begin{minipage}[b]{0.5\textwidth}
        \centering
        \includegraphics[height=5.705cm]{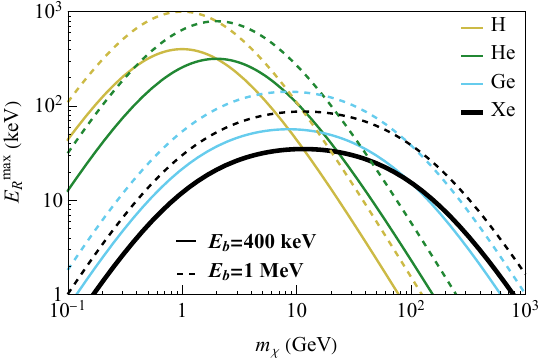} 
  \end{minipage}
  \begin{minipage}[b]{0.5\textwidth}
    \centering
    \includegraphics[height=5.705cm]{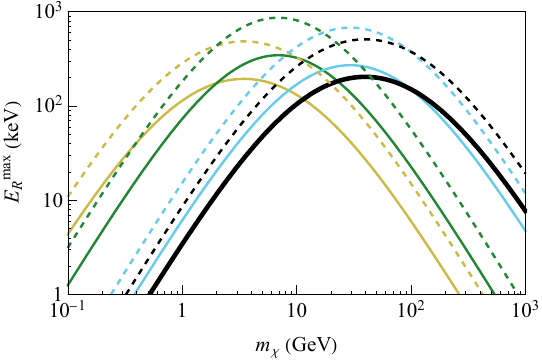}
  \end{minipage} 
\caption{Maximum nuclear target recoil energies $E_R^{\rm max}$ for dark matter upscattered by beams of protons~(left) or carbon~(right) with kinetic energies $E_b = 0.4\,\mev$~(solid) and $E_b = 1.0\,\mev$~(dashed) for a selection of target nuclei.}
\label{fig:kinematics}
\end{figure*}

In Fig.~\ref{fig:kinematics} we show the maximum nuclear recoil energies $E_R^{\rm max}$ setting $\cos\theta = 1$ for DM upscattered by beams of protons~(left) or carbon~(right) with kinetic energies $E_b = 0.4\,\mev$~(solid) and $E_b = 1.0\,\mev$~(dashed) on targets of hydrogen~(H), helium~(He), germanium~(Ge), and xenon~(Xe). These recoil energies fall within the sensitivity windows of typical underground nuclear recoil DM detectors.

Given an accelerator beam of particles with energy $E_b$, total current $I_b$, and charge per particle $Q_b$, the differential rate of DM upscattering per unit beam travel length is
\beq
\frac{dN_{\chi}}{dt\,dz\,dc_\theta} = \frac{I_b}{Q_b}\,n_\chi\,\frac{d\sigma_{\chi b}}{dc_\theta} \ ,
\eeq
where $c_\theta = \cos\theta$ corresponds to the outgoing DM angle relative to the beam, $z \in [-L/2,\,L/2]$ ranges over the beam travel region after full acceleration, and $n_\chi$ is the local $\chi$ DM number density. From this, we see that total rate of upscattered DM is proportional to the quantity 
\beq
\hspace{-0.35cm}\frac{I_b}{Q_b}L=6\times10^{17}~{\rm \frac{cm}{s}}\left(\frac{I_b}{1~\rm mA}\right)\left(\frac{Q_p}{Q_b}\right)\left(\frac{L}{100~\rm cm}\right)  ,
\label{eq:beamfluence}
\eeq
where $L$ is the total length over which the fully accelerated beam travels.

\subsection{Detection of upscattered dark matter}

For a detector placed downstream of the beam, the measured rate of DM scattering in the detector is
\begin{widetext}
\beq
R &=& \int_{-L/2}^{L/2}\!dz\int\!dc_\theta\;
\frac{dN_{\chi}}{dt\,dz\,dc_\theta}\,
\bigg(1-e^{-\ell\,\sigma_{\chi N}\,n_N}\bigg)
\,P_{\rm thr}(\theta;E_{\rm thr})\,P_{\rm sh}(\theta,z)
\label{eq:rate}\\
&=&
\frac{I_b}{Q_b}\,n_\chi\,\sigma_{\chi b}\,L\;\times
\int_{-1/2}^{1/2}\!d({z}/{L})\int\!dc_\theta\;
\frac{1}{\sigma_{\chi b}}\frac{d\sigma_{\chi b}}{dc_\theta}\,
\bigg(1-e^{-\ell\,\sigma_{\chi N}\,n_N}\bigg)
\,P_{\rm thr}(\theta;E_{\rm thr})\,P_{\rm sh}(\theta,z) \ ,
\nnmb
\eeq
\end{widetext}
where $\ell = \ell(\theta,z)$ is the path length in the detector for a DM particle upscattered at point $z$ through angle $\theta$, $n_N$ is the number density of the target nucleus, $\sigma_{\chi N}$ is the total DM-nucleus cross section, $P_{\rm thr}(\theta,E_{\rm thr})$ is the probability that the scattering will yield a recoil energy above the detector threshold $E_{\rm thr}$, and $P_{\rm sh}(\theta,z)$ is the probability for DM to scatter in material before reaching the detector. The exponential factor is the probability for a DM particle from $(z,\theta)$ to scatter at least once in the detector; it reduces to $\ell\,\sigma_{\chi N}\,n_N$ when this combination is much less than unity. In the second line of Eq.~\eqref{eq:rate}, we have factored the expression into a total upscattering rate times a dimensionless acceptance factor for scattering above threshold in the detector.

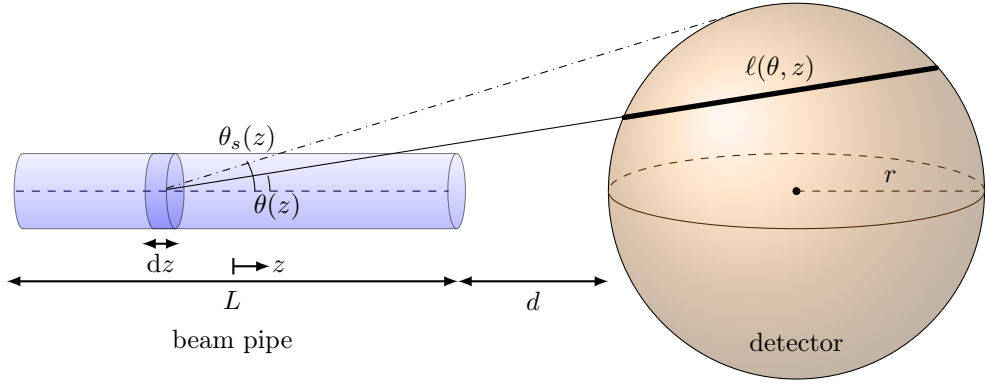
\begin{figure*}[ttt]
\centering
\begin{tikzpicture}
    
    \node [draw, shape=cylinder, minimum height=6cm, minimum width=1cm, cylinder end fill=blue, left color=blue!10!white, right color=blue!50!white, middle color=blue!30!white, shading angle=0,opacity=0.5] {};
    \node [draw, shape=cylinder, minimum height=0.5cm, minimum width=1cm, cylinder end fill=blue, left color=blue!20!white, right color=blue!60!white, middle color=blue!40!white, shading angle=0,opacity=0.5] at (-1,0) {};
    \node (left) at (-3,0) {};
    \node (right) at (3,0) {};
    \node (dz) at (-1,0) {};
    \node at (0,-2) {beam pipe};
    
    \shade[ball color = orange!70, opacity = 0.4] (7.5,0) circle (2.5cm);
    \draw (7.5,0) circle (2.5cm);
    \draw[orange!30!black] (5,0) arc (180:360:2.5 and 0.5);
    \draw[dashed,orange!30!black] (10,0) arc (0:180:2.5 and 0.5);
    \fill[fill=black] (7.5,0) circle (1.5pt);
    \draw[dashed,orange!30!black] (7.5,0) -- node[above,text=black]{$r$} (10,0);
    \node at (7.5,-2) {detector};
    
    \draw[dashed,blue!30!black] (left) -- (right);
    \draw[thick,latex-latex] (-1.2,-0.7) -- node[below] {$\dd{z}$} (-0.7,-0.7);
    \draw[thick,latex-latex] (-3,-1.2) -- node[below] {$L$} (3,-1.2);
    \draw[thick, |-latex] (0,-1) -- node[right=0.15cm] {$z$} (0.5,-1);
    
    \draw[thick,latex-latex] (3,-1.2) -- node[below] {$d$} (5,-1.2);
    \draw (0.5,0) arc (0:15:0.8);
    \node[below] at (0.6,0.1) {$\theta(z)$};
    \draw (dz) -- (5.2,0.99);
    \draw[line width=2pt,line cap=round] (5.21,0.98) -- node[above] {$\ell(\theta,z)$} (9.36,1.64);
    \draw[dash dot] (dz) -- (6.68,2.36);
    \draw (0.3,0) arc (0:30:0.8);
    \node[above] at (0.2,0.4) {$\theta_s(z)$};
    
    \end{tikzpicture}
\caption{Effective path length $\ell(\theta,z)$ in a spherical detector located along the beam axis.}
\label{fig:sphere}
\end{figure*}

The result of Eq.~\eqref{eq:rate} is very general, and it is instructive to evaluate its components for a specific detector geometry. We consider a spherical detector of radius $r$ located along the beam axis a distance $d$ from the end of the beam pipe, as illustrated in Fig.~\ref{fig:sphere}. For this configuration, the DM path length is
\beq
\ell(\theta,z) = \Theta(\theta_s-\theta)\;2\,r\,\sqrt{1-D^2\sin^2\theta/r^2} \ ,
\eeq
where $D = L/2 - z + d + r$ and $\theta_s = \sin^{-1}(r/D)$.

\subsection{Application to spin-independent pointlike DM}
\label{sec:saturation}
If we specialize further to DM that scatters primarily through a spin-independent point interaction, the differential DM-nucleus cross section is
\beq
\frac{d\sigma_{\chi N}}{dE_R} = \frac{1}{E_R^{\rm max}}\,|F_N(E_R)|^2\,\bar{\sigma}_{N} \ ,
\label{eq:dsigder}
\eeq
where $E_R \leq E_R^{\rm max} = 2\,\mu_{\chi N}^2v_\chi^2/m_N$, $F_N(E_R)$ is a nuclear form factor for SI scattering~~\cite{Engel:1992bf,Lewin:1995rx},
and $\bar{\sigma}_N$ is given in the Born approximation by
\beq
\bar{\sigma}_N = \lrf{\mu_{\chi N}}{\mu_{\chi p}}^2A^2\,\sigma_{\chi n} \ ,
\label{eq:pernuc}
\eeq
for an effective per-nucleon cross section $\sigma_{\chi n}$. 

Since we study very large cross sections in this work, we also consider the possibility that the Born approximation on which Eq.~\eqref{eq:pernuc} is based might break down. While the way in which this occurs depends on the detailed interactions between DM and nucleons, there exists a simple prescription based on geometric saturation that provides a reasonable approximation to calculations in a wide range of models~\cite{Pospelov:2019vuf,Digman:2019wdm}. Specifically, we bound from above the total nuclear cross section derived from Eq.~\eqref{eq:dsigder} by the geometric cross section $\sigma_{\chi N} \leq 4\pi r_N^2$ with $r_N \simeq 1.2\,\text{fm}\,A^{1/3}$. This is equivalent to the replacement of $\bar{\sigma}_N$ in Eq.~\eqref{eq:dsigder} by $\bar{\sigma}_{N,\rm eff}$ defined by
\beq
\bar{\sigma}_{N,\rm eff} = \left\{
\begin{array}{ccc}
\bar{\sigma}_N&~~;~~&\sigma_{\rm tot} < 4\pi\,r_N^2\\
&&\\
\frac{4\pi\,r_N^2}{ \int_0^1\!dx\;|F_N(xE_R^{\rm max})|^2}&;&\sigma_{\rm tot} > 4\pi\,r_N^2
\end{array}\right.
\label{eq:sigeff}
\eeq
where
\beq
\sigma_{\rm tot}  = \bar{\sigma}_N\,\int_0^1\!dx\;|F_N(xE_R^{\rm max})|^2 \ .
\eeq
With this form, we can express the nuclear cross section portion of Eq.~\eqref{eq:rate} by
\beq
\begin{aligned}
&\sigma_{\chi N}\,P_{\rm thr}(\theta;E_{\rm thr}) = 
\bar{\sigma}_{N,\rm eff}\,
\\
&\quad\quad\times\int_{x_{\rm thr}}^1\!dx\;|F_N(xE_R^{\rm max})|^2\;\Theta(1-x_{\rm thr}) \ ,
\end{aligned}
\eeq
where $x_{\rm thr} = E_{\rm thr}/E_R^{\rm max}$.

We can also specify the upscattering rate more precisely if we specialize to a SI interaction. For a low-energy beam of protons,
\beq
\frac{d\sigma_{\chi p}}{dc_\theta} = 2\,\sigma_{\chi p}\,\cos\theta \ .
\eeq
If the interaction connects DM to protons and neutrons with equal strength, we can identify $\sigma_{\chi p}= \sigma_{\chi n}$ defined in Eq.~\eqref{eq:pernuc}. This result can also be generalized to low-energy beams of nuclei. Using the saturation prescription described above, we find
\beq
\frac{d\sigma_{\chi b}}{dc_\theta} = 2\,\cos\theta\;\bar{\sigma}_{b,\rm eff}\,|F_b(E_{R,b})|^2 \ ,
\eeq
with
\beq
E_{R,b} = \frac{2\mu_{\chi b}^2}{m_b}
\lrf{2E_b}{m_b}\cos^2\theta \ , 
\eeq
and $\sigma_{b,\rm eff}$ defined as for $\bar{\sigma}_{N,\rm eff}$ in Eq.~\eqref{eq:sigeff} with $N\to b$, $E_R \to E_{R,b}$, and $E_R^{\rm max} \to E_{R,b}^{\rm max} = E_{R,b}(c_\theta = 1)$.

\begin{figure*}[ttt]
\centering
\includegraphics[height=5.2cm]{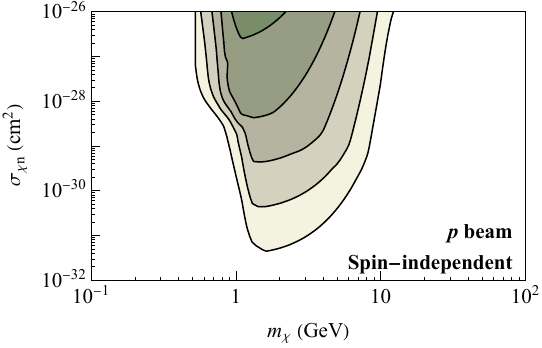} 
\includegraphics[height=5.2cm]{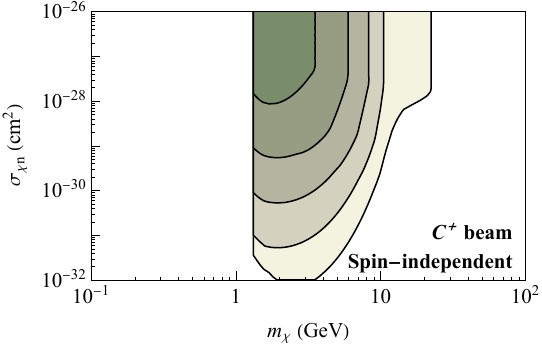}
\raisebox{0.13\height}{\includegraphics[height=4.6cm,trim=0.2cm 0 0 0,clip]{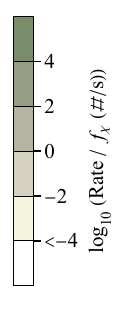}}

\caption{Dark matter scattering rates in a xenon detector located downstream of a beam section with $L = 5\,\text{m}$ and $E_b = 0.4\,\mev$ for beams of protons with $I_b = 1.0\,\text{mA}$~(left), and carbon~${}^{12}$C$^+$ with $I_b = 0.1\,\text{mA}$~(right). In both cases, the detector is assumed to be a sphere of radius $r= 10\,\text{cm}$ located a distance $d = 50\,\text{cm}$ from the end of the beam pipe with a lower energy threshold of $E_\text{thr}=5\,\kev$.}
\label{fig:ratesxe}
\end{figure*}

In Fig.~\ref{fig:ratesxe} we show the estimated detector rates of beam upscattered DM as a function of mass $m_\chi$ and per nucleon cross section $\sigma_{\chi n}$ assuming a pointlike SI interaction for a potential beam and detector apparatus. We take beam parameters motivated by the LUNA accelerator~\cite{FORMICOLA2003609,Costantini:2009wn} with an accelerated beam section of $L=5\,\text{m}$ and a kinetic energy per particle of $E_b=0.4\,\mev$ for proton beams with current $I_b = 1.0\,\text{mA}$~(left) and carbon $^{12}$C$^+$ beams with current $I_b=0.1\,\text{mA}$~(right). For both beam types, we assume a detector consisting of a sphere containing liquid xenon of radius $r = 10\,\text{cm}$ located along the beam axis at a distance $d=50\,\text{cm}$ from the end of the beam pipe with a lower detection energy threshold of $E_{\rm thr} =5\,\kev$. See Fig.~\ref{fig:sphere} for details of the setup.

The detector scattering rates shown in Fig.~\ref{fig:ratesxe} are significant and suggest that this method could be used to test strongly interacting DM even for fractional densities $f_\chi \ll 1$. These rates trace the DM density enhancements shown in Fig.~\ref{fig:ndm} to a large degree. They are largest for $m_\chi \sim 1$--$10\,\gev$, corresponding to the enhanced thermal DM population discussed in Sec.~\ref{sec:capture}, although there is also a shoulder at larger masses from the traffic jam population. For masses below $m_\chi \sim 1\,\gev$, the detection rates are reduced by the lower DM population due to evaporation as well as the energy threshold we assume for the detector. This is most clearly visible in the right panel of Fig.~\ref{fig:ratesxe} where the kinematic mismatch between the masses of the beam carbon nuclei and the $m_\chi \sim 1\,\gev$ DM leads to reduced upscattering energies and a sharp cutoff for DM masses below a GeV. We also note that the scattering rates fall off more quickly at smaller cross sections than the DM density. This arises from the dependence of the detector acceptance on the cross section since smaller cross sections produce a lower probability for the DM to scatter in the detector as it passes through.

\begin{figure*}[ttt]
\centering
\begin{minipage}[b]{0.5\textwidth}
        \centering
        \includegraphics[height=5.705cm]{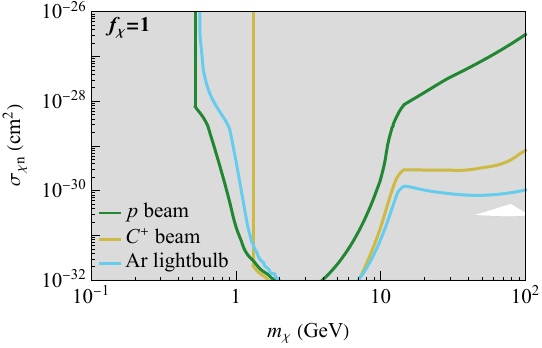} 
  \end{minipage}
  \begin{minipage}[b]{0.5\textwidth}
    \centering
    \includegraphics[height=5.705cm]{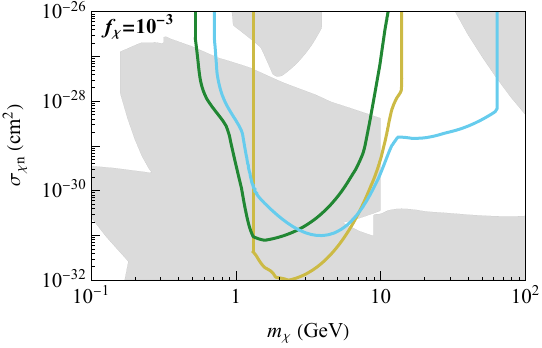}
  \end{minipage} 
\begin{minipage}[b]{0.5\textwidth}
        \centering
        \includegraphics[height=5.705cm]{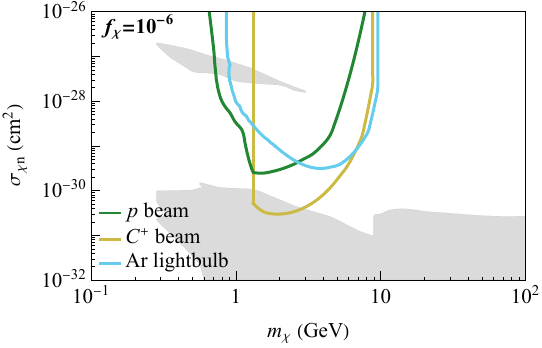} 
  \end{minipage}
  \begin{minipage}[b]{0.5\textwidth}
    \centering
    \includegraphics[height=5.705cm]{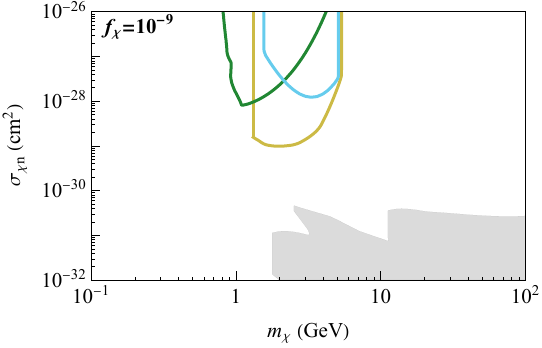}
  \end{minipage} 
\caption{Experimental reach of the nuclear accelerator setup for proton beams (green) and singly ionized carbon beams (yellow) with a 10-cm radius xenon detector, and of an argon thermal source with a 1 g gaseous carbon monoxide detector with a nuclear recoil threshold of 0.8 eV. The contour lines are for 10 events per year and a spin-independent interaction. The shaded gray regions show the combined exclusions on this form of dark matter component and summarize the results of Appendix~\ref{sec:appb} presented in Fig.~\ref{fig:SIDMbounds}. 
}
\label{fig:results}
\end{figure*}

In Fig.~\ref{fig:results}, we compare the potential reach of the accelerator upscattering method to existing bounds on DM with SI scattering on nuclei for DM fractions $f_{\chi} = 1,\,10^{-3},\,10^{-6},\,10^{-9}$ as a function of the DM mass $m_\chi$ and per nucleon cross section $\sigma_{\chi n}$. The green~(proton beam) and yellow~(C$^+$ beam) contours encircle the regions that would yield at least 10 events per year for beam and (liquid xenon) detector parameters as described above for Fig.~\ref{fig:ratesxe}. Also shown in the figure are blue contours for the corresponding reach for the thermal upscattering method to be discussed in Sec.~\ref{sec:hot}. The gray shaded regions show the combination of current exclusions on such a DM candidate as detailed in Fig.~\ref{fig:SIDMbounds} and discussed in Appendix~\ref{sec:appb}. These plots show that accelerator upscattering can potentially test fractional DM components beyond what has been achieved so far. 

While our analysis neglects several potentially important aspects such as backgrounds and attenuation of upscattered DM due to shielding,
we argue that this technique is a promising and realistic method for testing strongly interacting DM. The detector configuration studied here is relatively modest and in line with smaller liquid xenon detectors such as the XENON10~\cite{XENON:2007uwm} and ZEPLIN-III~\cite{ZEPLIN-III:2009htd} experiments. Moreover, these experiments achieved near-zero backgrounds, albeit within less active environments. 

A potential source of backgrounds relative to these more traditional dark matter search experiments are those from the beam itself as well as the nuclear targets the beam is directed to. Some of these can be mitigated or avoided by choosing a suitable detector location. For LUNA-MV, the primary central beam is split and directed to a pair of target stations located at large angles~\cite{SEN2019390,Prati:2020uxj}. The current setup would allow a detector to be seated directly behind the beam splitter and away from the beam path and target stations. Similarly, JUNA~\cite{juna} and CASPAR~\cite{caspar} feature a bend in the beam after the initial accelerating segment and a section of evacuated beam pipe, and could similarly accommodate a detector directly behind the bending magnets. Residual beam backgrounds produced with the low beam energies of these accelerators have very small penetrating power in the case of protons and charged nuclei, while secondary neutrons and photons may require a layer of high $Z$ material to mitigate. Much thicker shielding placed around the detector away from a beam aperture would reduce backgrounds from the nuclear reactions studied in the target stations, although it may be necessary to avoid searching for DM while neutron-producing reactions are being studied. These background sources would be relatively straightforward to measure \textit{in situ} before staging a DM search.

For very large DM cross sections, the upscattered DM beam can be attenuated by accelerator components and detector shielding between the beam pipe and the detector material. We estimate that the attenuation of DM from a reasonable thickness $\ell < 5\,\text{cm}$ of steel or lead is minimal but can be significant for thicknesses greater than this. To avoid losing potential signal to shielding, a mostly unshielded beam aperture can be left for upscattered DM to pass through combined with thicker shielding elsewhere.

Despite these important practical challenges, the large DM rates we find suggest that they could be overcome. Of course, a more detailed study beyond the scope of the present work would be needed before installing an actual detector.

\section{Upscattering of Dark Matter by Thermal Sources\label{sec:hot}}

We seek next to determine the upscattering rate of strongly interacting dark matter using very hot thermal sources. This second, complimentary method is presented in the context of hot gas surrounding the filament of a conventional incandescent lightbulb. Given that such filaments can heat gas to temperatures on the order of $\mathcal{O}(10^3\,\text{K})$, DM could be scattered to kinetic energies near $\mathcal{O}(\text{eV})$ by the high energy Boltzmann tail. Such energies are large enough to be potentially detectable at near-future experimental setups.

\subsection{Thermal upscattering of DM}

Consider a gas of SM particles heated to a temperature $T_{g}$. This gas can scatter on a population of DM thermalized with the Earth at a temperature $T_\chi\ll T_{g}$, accelerating the DM and potentially making it more easily detectable. For example, the filaments in typical commercial lightbulbs are heated to around $3000~{\rm K}$, along with a substantial number of gas particles surrounding the filament. To estimate this number, we assume that the volume occupied by these hot gas particles is $\pi\ell_{\rm coll}^2\ell_{\rm fil}$ with $\ell_{\rm coll}$ the scattering length of the gas and $\ell_{\rm fil}$ the length of the filament. Taking $\ell_{\rm coll}=0.5~\rm mm$, $\ell_{\rm fil}=20~\rm cm$, and a gas pressure of $0.7~\rm atm$ leads to the crude estimate of $N_{g}\sim 10^{17}$ hot gas particles. In this case the rate of hot gas particles a unit length of DM particles encounters is 
\begin{equation}
\begin{aligned}
N_{g}v_{g}&\sim N_{g}\sqrt{\frac{T_{g}}{m_{g}}}
\nonumber\\
&\sim 10^{22}~\frac{\rm cm}{\rm s}\left(\frac{N_{g}}{10^{17}}\right)\left(\frac{T_{g}}{3000~\rm K}\right)^{1/2}\left(\frac{m_{^4{\rm He}}}{m_{g}}\right)^{1/2},
\label{eq:LBfluence}
\end{aligned}
\end{equation}

where we have normalized the mass of the gas particles, $m_{g}$, to that of $^4{\rm He}$. Comparing this to Eq.~(\ref{eq:beamfluence}), we see that it corresponds to a substantially larger rate of DM upscattering, albeit at much lower energies.

Assuming that the gas nuclei scatter on DM through a spin- and momentum-independent interaction with cross section $\sigma_{\chi g}$, the flux of DM a distance $d$ from the region of hot gas is
\begin{equation}
\begin{aligned}
\Phi_\chi&=\frac{N_{g}v_{g}\sigma_{\chi g}n_\chi}{4\pi d^2}
\nonumber\\
&\sim \frac{2\times 10^4}{\rm cm^2~s}\left(\frac{N_{g}}{10^{17}}\right)\left(\frac{T_{g}}{3000~\rm K}\right)^{1/2}\left(\frac{m_{^4{\rm He}}}{m_{g}}\right)^{1/2}
\nonumber\\
&\quad\quad\times\left(\frac{A_{g}}{4}\right)^4\left(\frac{\sigma_{\chi n}}{10^{-27}~\rm cm^2}\right)\left(\frac{n_\chi}{10^9~\rm cm^{-3}}\right)\left(\frac{10~\rm cm}{d}\right)^2.
\label{eq:lbflux}
\end{aligned}
\end{equation}
In this expression we have assumed that the cross section for DM to scatter on the gas nuclei scales as $A_{g}^4$ and has not reached geometric saturation. The kinetic energy distribution of the upscattered DM, in the limit that $T_\chi \ll T_{g}$ and assuming that the interaction matrix element is momentum independent, is\footnote{We show the expression for finite $T_\chi$ in Appendix~\ref{sec:appa}.}
\begin{equation}
\begin{aligned}
\frac{1}{\Phi_\chi}\frac{d\Phi_\chi}{dE_\chi}&\simeq\frac{m_{g}m_\chi}{4\mu_{\chi g}^2 T_{g}}\exp\left(-\frac{m_{g}m_\chi}{4\mu_{\chi g}^2 T_{g}}E_\chi\right).
\end{aligned}
\label{eq:dm-lbdist}
\end{equation}
$\mu_{\chi g}$ is the reduced mass of the DM and gas. Thus, the typical kinetic energy of DM upscattered by the hot gas is of order $T_{g}$ unless there is a large hierarchy between $m_\chi$ and $m_{g}$. With $T_{g}=3000~{\rm K}=0.26~{\rm eV}$, a detector with an energy threshold of order $0.1$ to $1~\rm eV$ placed near the hot gas can potentially detect the upscattered DM.

\subsection{Detection of thermally upscattered DM}

Gas-based detectors may offer the best prospects for sensitivity to eV-scale energy depositions from nuclear scattering in the near future. NEWS-G~\cite{Giomataris:2008ap,*Savvidis:2016wei,*NEWS-G:2017pxg,*Arnaud:2018bpc} has demonstrated energy thresholds on the order of hundreds of eV. Pushing even lower, exciting rotational and vibrational molecular modes in a gas could offer a pathway to detecting ${\cal O}(0.1~{\rm eV})$ energy depositions~\cite{Essig:2019kfe}. In the following, we roughly estimate the sensitivity of an idealized future detector, assuming that the detector is sensitive to all DM with kinetic energy above $E_\chi^{\rm min}$ with a probability of interacting given by $L_{\rm det}/\ell_{\rm scatt}$ where $L_{\rm det}$ is the length of the detector and $\ell_{\rm scatt}$ is the DM scattering length in the detector material. We leave estimates folding in realistic detection efficiencies for future work.

We will crudely assume that the scattering length in the detector is dominated by elastic scattering of DM on detector nuclei with a cross section $\sigma_{\chi N}$ so that $\ell_{\rm scatt}\propto 1/\sigma_{\chi N}$. The rate of scattering near a thermal source producing the flux of DM given in Eq.~(\ref{eq:lbflux}) is then
\begin{equation}
\begin{aligned}
R&\sim\Phi_\chi N_{\rm det}\sigma_{\chi N}f_{\rm det}\sim \frac{60}{\rm gram~s}\left(\frac{N_{g}}{10^{17}}\right)\left(\frac{T_{g}}{3000~\rm K}\right)^{1/2}
\\
&\quad\quad\times\left(\frac{m_{^4{\rm He}}}{m_{g}}\right)^{1/2}\left(\frac{A_{g}}{4}\right)^4\left(\frac{A_{\rm det}}{4}\right)^4\left(\frac{\sigma_{\chi n}}{10^{-27}~\rm cm^2}\right)^2
\\
&\quad\quad\times\left(\frac{n_\chi}{10^9~\rm cm^{-3}}\right)\left(\frac{10~\rm cm}{d}\right)^2\left(\frac{f_{\rm det}}{0.1}\right).
\label{eq:rateLBdet}
\end{aligned}
\end{equation}
In this expression, $N_{\rm det}$ is the number of nuclei in the detector with atomic number $A_{\rm det}$, and $f_{\rm det}$ is the fraction of DM with kinetic energy above $E_\chi^{\rm min}$ that can be detected,
\begin{equation}
\begin{aligned}
f_{\rm det}&=\frac{1}{\Phi_\chi}\int_{E_\chi^{\rm min}}^\infty dE_\chi \frac{d\Phi_\chi}{dE_\chi}\simeq\exp\left(-\frac{m_{g}m_\chi}{4\mu_{\chi g}^2 T_{g}}E_\chi^{\rm min}\right).
\end{aligned}
\end{equation}
This approximation is valid when the thermalized DM temperature can be neglected, $T_\chi\ll T_g$. The numerical estimate in Eq.~(\ref{eq:rateLBdet}) uses the scaling $\sigma_{\chi N}\propto A_{\rm det}^4\sigma_{\chi n}$ to relate the cross section to scatter on a nucleus to that on a nucleon, assuming a point interaction with neutrons and protons of equal strength. In the results that follow, however, we make use of the more sophisticated relation between $\sigma_{\chi N}$ and $\sigma_{\chi n}$ described in Sec.~\ref{sec:saturation}.

\begin{figure*}[ttt]
\centering
\includegraphics[height=5.2cm]{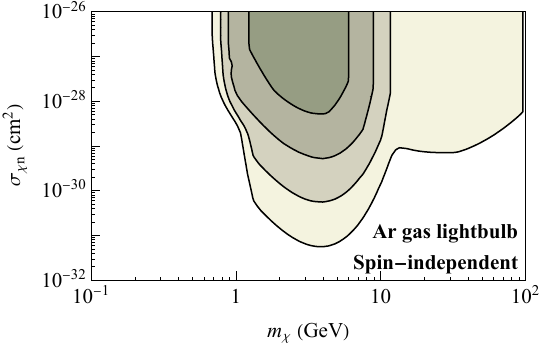} 
\includegraphics[height=5.2cm]{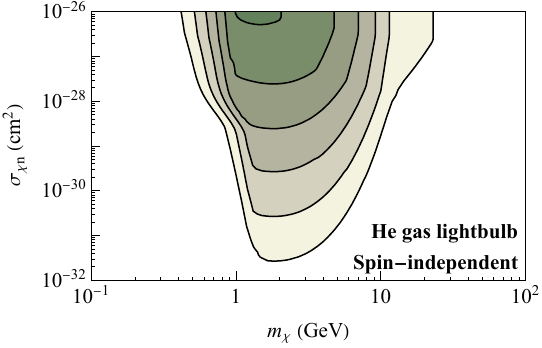}
\raisebox{0.13\height}{\includegraphics[height=4.6cm,trim=0.2cm 0 0 0,clip]{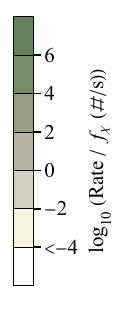}}

\caption{Dark matter scattering rates in a 1-g ${\rm CO}$ detector with a lower energy threshold of $E_\chi^\text{min}=0.8$ eV located a distance $d=10$ cm from a thermal source with $10^{17}$ ${\rm Ar}$ atoms (left) and ${\rm He}$ atoms (right) at $3000~\rm K$.}
\label{fig:bulbrates}
\end{figure*}

In Fig.~\ref{fig:bulbrates}, we show the per-gram detector rates as a function of mass $m_\chi$ and per-nucleon cross section, assuming a pointlike SI interaction using the DM densities computed in Sec~\ref{sec:capture}. We fix $N_g=10^{17}$ with $T_{g}=3000~K$, a distance $d=10$~cm away from the detector which should allow for sufficient thermal insulation. We assume that the hot gas is helium or argon and that the accelerated DM scatters on carbon nuclei in a detector filled with carbon monoxide gas. Optimistically, we set $E_\chi^{\rm min}=0.8~\rm eV$; detailed studies to more accurately model the detector response are beyond the scope of this work.

Another possibility for low-threshold detectors that has been extensively discussed in the DM direct detection literature involves DM scattering on electrons~\cite{Essig:2011nj}. Unfortunately, this does not obviously provide a clear pathway for the detection of DM upscattered by nuclei in a hot gas. This is largely due to the fact that DM that is most efficiently captured in the Earth and subsequently upscattered by nuclei has $m_\chi \sim 1$--$100\,\gev$ (cf. Fig.~\ref{fig:ndm}) which is not well kinematically matched to maximize the momentum transfer when scattering on electrons.

\section{Realizations of Strongly Interacting Dark Matter\label{sec:models}}

Having presented two promising search techniques for a strongly interacting dark matter component $\chi$ with mass in the range $m_\chi \sim 1$--$100\,\gev$, we turn next to an examination of potential candidates for such a species. We focus on two possibilities: DM made up partially or entirely from particles charged under the strong force; and DM connecting to the SM through a new dark mediator particle. As a specific realization of the second class of candidates, we perform a detailed study of asymmetric DM coupled to a dark photon.

\subsection{Dark matter with QCD interactions}

Proposals for stable relics with constituents charged under quantum chromodynamics~(QCD) have a long history as well as more recent interest. Early examples include stable di-baryons consisting of six quarks~\cite{Jaffe:1976yi} and tightly bound quark nuggets of macroscopic size~\cite{Witten:1984rs}. Recent studies of DM connected to QCD include candidates composed of the standard quarks and gluons, both microscopic~\cite{Jaffe:1976yi,Farrar:2003gh,Farrar:2005zd,Farrar:2017eqq} and macroscopic~\cite{Witten:1984rs,Oaknin:2003uv,Bai:2018vik}, as well as DM candidates containing exotic particles charged under the strong force~\cite{DeLuca:2018mzn,Beylin:2020bsz}. Among these proposals, the methods presented in this work are most applicable to microscopic relics consisting only of quarks and gluons. They could also be applied to bound states involving new QCD-charged states, but these are very strongly constrained by direct~\cite{CMS:2017abv,ATLAS:2017mjy,Evans:2018scg} and indirect~\cite{Becciolini:2014lya,Llorente:2018wup} searches at the Large Hadron Collider~(LHC) and must be heavier than at least a few hundred GeV to have avoided detection, putting them beyond our primary region of sensitivity. 

The best-studied proposals for a stable QCD relic are the sexaquark~(S) and the H-dibaryon with quark quantum numbers $uuddss$~\cite{Jaffe:1976yi,Farrar:2003gh,Farrar:2005zd,Farrar:2017eqq}. The state S must lie in the mass range $m_S \sim 1860$--$1890\,\mev$ to be stable while also not destabilizing nuclei~\cite{Farrar:2017eqq,Farrar:2018hac,Kolb:2018bxv}, while the H-dibaryon would be slightly heavier at around $2150$ MeV~\cite{Jaffe:1976yi}. Direct searches for a stable di-baryon have been inconclusive~\cite{Belle:2013sba,ALICE:2018ysd} and its study on the lattice is very challenging~\cite{Green:2021qol}. There has also been some debate over the astronomical properties of a stable di-baryon, with Ref.~\cite{Kolb:2018bxv} finding that it would obtain a very small density fraction $n_S/n_{\rm DM} \sim 10^{-11}$ through thermal freeze-out, and Ref.~\cite{McDermott:2018ofd} arguing that the existence of this state would drastically change the evolution and properties of neutron stars. Both of these works rely on reasonable estimates for di-baryon interaction cross sections, while Refs.~\cite{Farrar:2020zeo,Farrar:2022mih,Shahrbaf:2022upc} counter that these can be significantly suppressed or avoided. We do not attempt to resolve this debate here, but we do note that a stable QCD relic, possibly making up only a small fraction of the total DM abundance, would be expected to have a mass and nucleon interaction cross sections within the range our proposals are most sensitive to.

\subsection{Asymmetric dark matter coupled to a dark photon}

Dark matter can also have a large nucleon scattering cross section if it connects to the SM through a relatively light mediator particle. In general, both lighter mediators and larger couplings increase the resulting cross section in the limit of zero momentum transfer. At the same time, these properties also tend to make the mediator easier to observe, and there is a generic tension between large DM-nucleon cross sections and direct bounds on the mediators~\cite{Knapen:2017xzo,Digman:2019wdm,Elor:2021swj}. 

As a concrete example of a DM candidate with a large nucleon cross section that also illustrates the tension with mediator searches, we investigate a Dirac fermion $\chi$ coupled to a massive dark photon that connects to the SM through gauge kinetic mixing. A similar analysis could be applied to other dark vectors that couple to baryons such as $B$~\cite{Tulin:2014tya,Gan:2020aco} or $B\!-\!L$~\cite{Heeck:2014zfa,Bauer:2018onh,Ilten:2018crw}. We focus on an asymmetric density of the dark relic $\chi$ to maintain a large enhanced density in the Earth.

The simple theory we consider consists of Dirac fermion $\chi$ coupled to a sub-GeV massive dark photon $A'$ with strength $g'$~\cite{Pospelov:2007mp,Pospelov:2008jd,ArkaniHamed:2008qn}. The corresponding Lagrangian is
\beq
\begin{aligned}
\mathscr{L} &\supset -\frac{1}{4}F_{\mu\nu}^\prime F^{\prime \mu\nu} 
+\frac{1}{2}m_{A'}^2A_{\mu}^{\prime}A^{\prime\mu}
\\
&+ \bar{\chi}i\gamma^{\mu}(\del_{\mu}+ig'A_{\mu}^{\prime})\chi
-m_\chi\bar{\chi}\chi
-\frac{\epsilon}{2\cos\theta_W}B_{\mu\nu}F^{\prime \mu\nu} \ ,
\label{eq:ldp}
\end{aligned}
\eeq
where the last term is the kinetic mixing with SM hypercharge. The dark photon mass can arise from a Higgs~\cite{Pospelov:2007mp,Pospelov:2008jd,ArkaniHamed:2008qn} or Stueckelberg~\cite{Stueckelberg:1957zz,Kors:2004dx} mechanism. For $m_{\chi} \sim 1$--$100\,\gev$, $m_{A'} \sim 10$--$100\,\mev$, and $\epsilon$ within the currently allowed range, we find that small relic fractions $f_{\chi} \ll 1$ and large per-nucleon cross sections $\sigma_{\chi n} \sim 10^{-32}$--$10^{-27}\,\text{cm}^2$ are obtained. In this range, we show that accelerator upscattering can test the theory beyond what has been possible through other methods, particularly when the relic density of $\chi$ is set in part by a charge asymmetry relative to $\bar{\chi}$.

Annihilation of the $\chi$ fermion in the early Universe is dominated by $\chi\bar{\chi} \to A'A'$ for $m_\chi > m_{A'}$ and $\alpha' = {g'}^2/4\pi \gg \epsilon^2\alpha$~\cite{Pospelov:2007mp}. In this limit, the Born-level annihilation cross section at zero velocity is
\beq
(\sigma v)_0 = 
\frac{\pi\,\alpha^{\prime 2}}{m_{\chi}^2}
\sqrt{1-\frac{m_{A'}^2}{m_{\chi}^2}} \ .
\label{eq:sigv0}
\eeq
For larger values of $\alpha^{\prime}$, the total cross section during annihilation in the early Universe can be enhanced by the Sommerfeld effect~\cite{Sommerfeld:1931qaf,Hisano:2006nn,ArkaniHamed:2008qn,Feng:2010zp} and bound state formation~\cite{Pospelov:2008jd,March-Russell:2008klu,Feng:2009mn,vonHarling:2014kha}. We follow the methods of Ref.~\cite{Baldes:2017gzw} to compute the relic density of $\chi$ while taking these effects into account and allowing for a possible charge asymmetry. In this approach, the mass of the $A^\prime$ mediator is neglected, which we have checked to be a good approximation during freeze-out for the parameter range we study by comparing the massless mediator result to an estimate for the massive result based on analytic results for the Hulth\`en potential~\cite{Cassel:2009wt,Hannestad:2010zt}. The resulting total relic fraction of $\chi$ and $\bar{\chi}$ can be written in the form~\cite{Graesser:2011wi}
\beq
f_\chi = \tilde{f}_{\chi}\,\lrf{1+r}{1-r} \ ,
\label{eq:dpdmfchi}
\eeq
where $r = n_{\bar{\chi}}/n_{\chi}$ is the ratio of $\bar{\chi}$ to $\chi$ relic densities, and $\tilde{f}_{\chi}$ is the DM fraction in the completely asymmetric limit ($r\to 0$) as determined by the charge asymmetry whose origin we do not specify. The ratio $r$ is obtained from a freeze-out calculation and decreases exponentially with an increasing annihilation cross section~\cite{Graesser:2011wi}.

The kinetic mixing interaction of Eq.~\eqref{eq:ldp} leads to SI scattering between relic $\chi$ particles and nucleons mediated by the dark photon. For $m_{A'} \ll m_Z$, the scattering is almost entirely with protons with a zero-momentum cross section of~\cite{Pospelov:2007mp}
\beq
\sigma_{\chi p} = 16\pi\,\epsilon^2\,\alpha\,\alpha'\,\frac{\mu_{p}^2}{m_{A'}^4}
\label{eq:sigpdp} \ .
\eeq
Extending this to scattering on nuclei $N=(A,Z)$ at nonzero momentum transfer $q = \sqrt{2m_N\,E_R}$, we have
\beq
\frac{d\sigma_{\chi N}}{dE_R} = 
\frac{1}{E_R^{\rm max}}\,
\sigma_{N}\,
|F_N(E_R)|^2
|F_{\chi}(E_R)|^2 \ ,
\label{eq:signdp}
\eeq 
where $\sigma_{\chi N} = (\mu_N/\mu_p)^2Z^2\sigma_p$, $F_N$ is the same nuclear form factor as before, and we have introduced a DM form factor
\beq
F_{\chi}(E_R) = \frac{1}{1+ q^2/m_{A'}^2} \ .
\label{eq:ffdp}
\eeq
It is straightforward to generalize our prescription for saturation of the total DM-nuclear cross section at $\sigma_{\chi N} \leq 4\pi\,r_N^2$ by replacing $|F_N|^2 \to |F_N|^2|F_\chi|^2$ in Eq.~\eqref{eq:sigeff}.

\begin{figure*}[ttt]
\begin{minipage}[c]{0.45\textwidth}
        \centering
        \includegraphics[width=\textwidth]{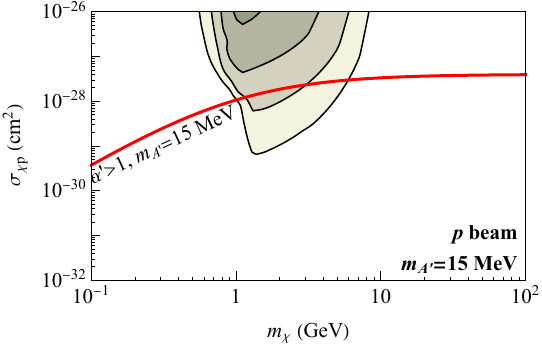} \\
        \includegraphics[width=\textwidth]{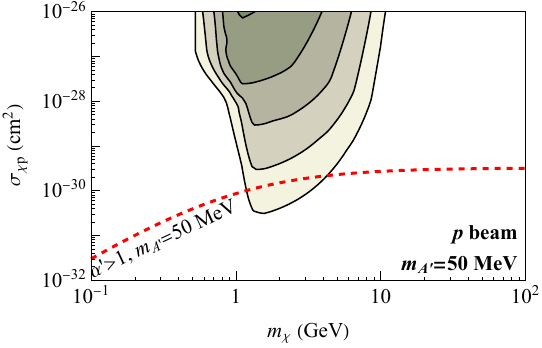} 
  \end{minipage}
  \begin{minipage}[c]{0.45\textwidth}
    \centering
    \includegraphics[width=\textwidth]{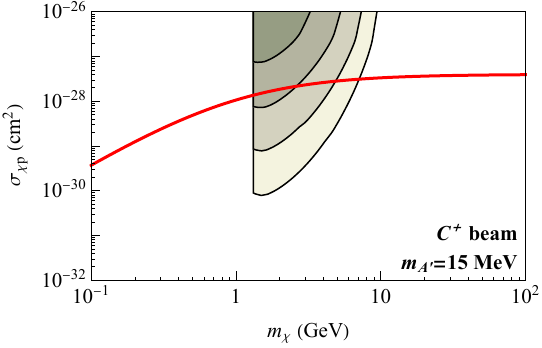} \\
    \includegraphics[width=\textwidth]{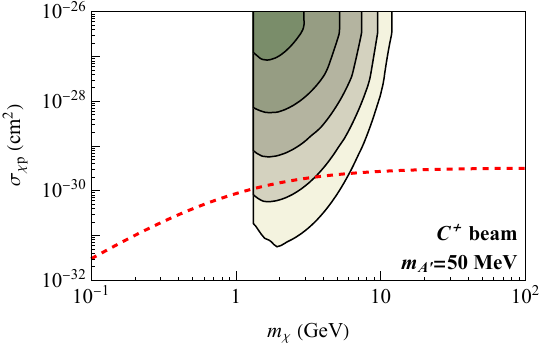}
  \end{minipage}
  \begin{minipage}[c]{0.1\textwidth}
    \centering
    \includegraphics[height=7cm]{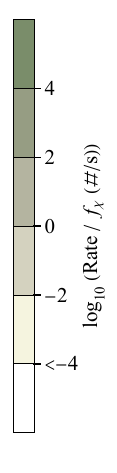}
  \end{minipage}

\caption{Dark photon-mediated dark matter scattering rates in a xenon detector located downstream of a beam section with $L = 5\,\text{m}$ and energy $E_b = 0.4\,\mev$ for beams of protons~(left) with $I_b = 1.0\,\text{mA}$, and carbon~${}^{12}$C$^+$~(right) with $I_b = 0.1\,\text{mA}$, for dark photon mediator masses of $m_{A^\prime} = 15\,\mev$~(top) and $m_{A^{\prime}} = 50\,\mev$~(bottom). In both cases, the detector is assumed to be a sphere of xenon of radius $r= 10\,\text{cm}$ located a distance $d = 50\,\text{cm}$ from the end of the beam pipe with a lower energy threshold of $E_\text{thr}=5\,\kev$. The solid (dashed) red lines show the boundary above which $\alpha'>1$ is needed to achieve the corresponding proton cross section for a 15 (50) MeV dark photon.}
\label{fig:ratesxedp}
\end{figure*}

In Fig.~\ref{fig:ratesxedp}, we show detector event rates for accelerator upscattered DM in this theory for benchmark dark photon masses of $m_{A'} =15\,\mev$~(top) and $m_{A'} = 50\,\mev$~(bottom). We assume the same accelerator and xenon detector parameters as in Sec.~\ref{sec:acc}, and we compute rates for a beam of energy of $E_b=0.4\,\text{m}$ and currents $I_b = 1.0\,\text{mA}$ for protons~(left) and $I_b=0.1\,\text{mA}$ for $^{12}$C$^{+}$~(right). For two reasons, the rates in this scenario are reduced relative to the SI point interaction considered in Sec.~\ref{sec:acc} and shown in Fig.~\ref{fig:ratesxe}. First, the nuclear cross section scales as $Z^2$ instead of $A^2$ prior to saturation. And second, for $m_{A'}=15\,\mev$, there is a significant additional suppression from the DM form factor of Eq.~\eqref{eq:ffdp} corresponding to $m_{A'}^2 < |q^2|$ in the beam or detector scattering. For $m_{A'}=50\,\mev$, we find that the DM form factor is typically close to unity.

\begin{figure*}[ttt]
\centering
\includegraphics[height=5.2cm]{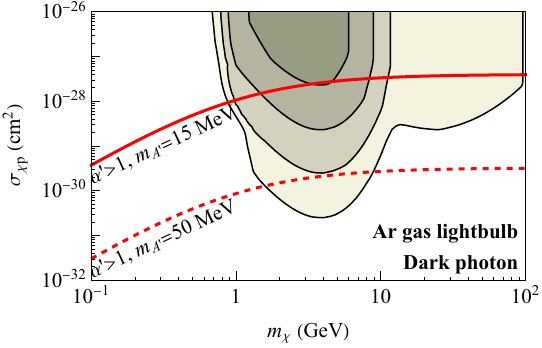}
\includegraphics[height=5.2cm]{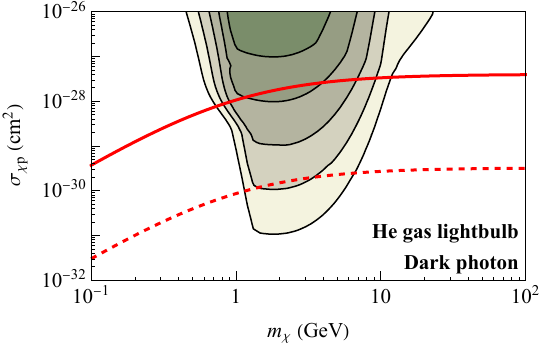}
\raisebox{0.13\height}{\includegraphics[height=4.6cm,trim=0.2cm 0 0 0,clip]{figs/legend4.pdf}}

\caption{Dark photon-mediated dark matter scattering rates in a 1\,g CO detector with a lower dark matter energy sensitivity of $E_\chi^\text{min}=0.8$ eV located a distance $d=10$ cm from a thermal source with $10^{17}$ ${\rm Ar}$ atoms~(left) or ${\rm He}$ atoms~(right) at $3000~\rm K$. The red solid (dashed) lines show the boundary above which $\alpha'>1$ is needed to achieve the corresponding proton cross section for a dark photon mediator mass of $m_{A'}=15$ MeV ($m_{A'}=50$ MeV).}
\label{fig:bulbratesdp}
\end{figure*}

In Fig.~\ref{fig:bulbratesdp}, we show the event rates per gram of gaseous CO detector for dark photon DM upscattered by a thermal source. As in Sec.~\ref{sec:hot}, we assume a source at temperature $T=3000\,\textrm{K}$ consisting of $10^{17}$ atoms, either argon~(left) or helium~(right). The detector is taken to consist of $1\,\text{g}$ of CO gas that is sensitive to DM kinetic energies above $E_{\chi}^\text{min} = 0.8\,\text{eV}$. As for accelerator upscattered DM, these rates are reduced relative to SI DM with equal couplings to protons and neutron due to the scaling with $Z^2$ rather than $A^2$ prior to (nuclear scattering saturation). In contrast, however, this method is not impacted by the benchmark dark photon masses ($m_{A'}=15,\,50\,\mev$) since the momentum transfers in this thermal source method are always much smaller, $q^2\ll m_{A'}^2$.

By working within a specific theory, we are able to derive consistency conditions on the range of nucleon scattering cross sections $\sigma_{\chi p}$. Fixing $m_{A'}$ and demanding $\alpha' \leq 1$ to maintain perturbativity, the largest possible $\sigma_{\chi p}$ corresponds to the greatest value of $\epsilon$ allowed by searches for dark photons~\cite{darkcast}. Assuming the dark photon decays visibly, the limit is approximately $\epsilon \lesssim 8\times 10^{-4}$ from NA48/2~\cite{NA482:2015wmo} and BaBar~\cite{BaBar:2014zli} for both $m_A'$ values. This upper bound on $\sigma_{\chi p}$ is shown by the solid red lines in Fig.~\ref{fig:ratesxedp} and agrees with Refs.~\cite{Knapen:2017xzo,Digman:2019wdm,Elor:2021swj}. A lower bound on $\sigma_{\chi p}$ can also be derived if we make the further assumption of a standard cosmological history and demand that the $\chi$ relic density be primarily asymmetric, which implies a lower bound on $\alpha'$ for a given DM fraction $f_\chi$. Combined with experimental lower bounds on $\epsilon$~\cite{NA64:2019auh,Blumlein:1990ay,Blumlein:1991xh,Blumlein:2011mv}, this provides a lower limit on the proton cross section given $m_{A'}$ and $f_{\chi}$. We find that the condition is not relevant to the parameter space tested by our methods.

\begin{figure*}[ttt]
\centering
\begin{minipage}[b]{0.5\textwidth}
        \centering
        \includegraphics[height=5.705cm]{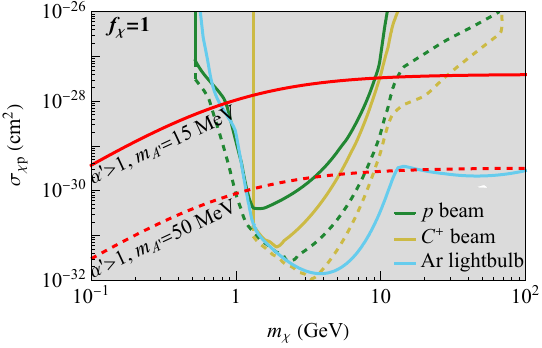} 
  \end{minipage}
  \begin{minipage}[b]{0.5\textwidth}
    \centering
    \includegraphics[height=5.705cm]{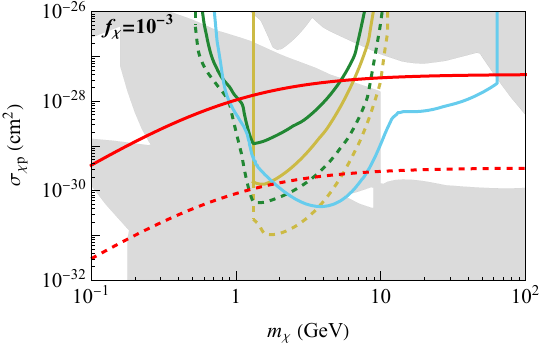}
  \end{minipage} 
\begin{minipage}[b]{0.5\textwidth}
        \centering
        \includegraphics[height=5.705cm]{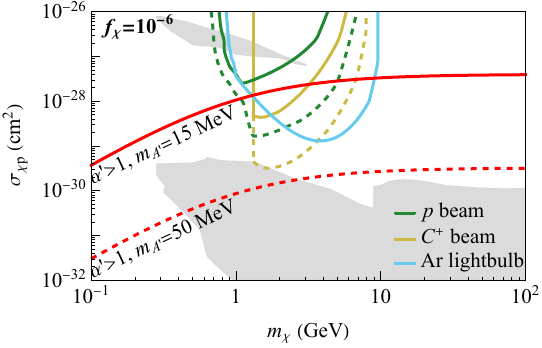} 
  \end{minipage}
  \begin{minipage}[b]{0.5\textwidth}
    \centering
    \includegraphics[height=5.705cm]{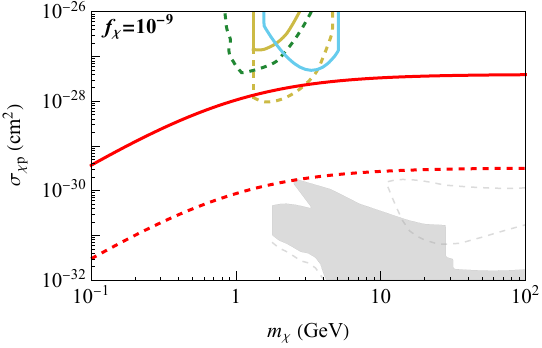}
  \end{minipage} 
    \caption{Experimental reach for dark photon dark matter of the nuclear accelerator setup with proton beams~(green) or singly ionized carbon beams (yellow) with a 10-cm radius xenon detector and for the argon lightbulb (blue) with a 1-g carbon monoxide detector of nuclear threshold 0.8 eV. The contour lines are for 10 events per year and a dark photon mediator of mass 15 MeV~(solid) or 50 MeV (dashed). The solid~(dashed) red line is the boundary above which $\alpha'>1$ is needed to achieve the corresponding proton cross section with a 15~(50) MeV dark photon. The gray shaded regions show current exclusions on this form of dark matter as discussed in Appendix~\ref{sec:appb} and shown in Fig.~\ref{fig:DPexclusions}, the dashed lines representing the additional exclusions for a 50 MeV dark photon.}
\label{fig:DPplots}
\end{figure*}

In Fig.~\ref{fig:DPplots}, we compare the sensitivity of our proposed accelerator upscattering method to this DM candidate to other bounds on it from direct detection and cosmology. The accelerator and detector parameters are the same as above. Each of the four panels in the figure correspond to different values of the $\chi$ DM fraction $f_{\chi} = 1,\,10^{-3},\,10^{-6},\,10^{-9}$. The contour lines in Fig.~\ref{fig:DPplots} correspond to event rates of at least 10 per year for proton~(p, green) or carbon (C$^+$, yellow) beams, or a lightbulb filled with argon gas (blue), and a dark photon mass of $15\,\mev$~(solid) or $50\,\mev$~(dashed). The gray regions show the collected exclusions on $\chi$ discussed in Appendix~\ref{sec:appb}.

\section{Conclusions\label{sec:conc}}

In this paper, we have proposed two new methods to detect strongly interacting dark matter (DM) that has accumulated in the Earth. Both methods leverage the tendency of strongly interacting DM in the local halo to scatter with nuclei in the Earth to slow down and thermalize. On one hand, this leads to a DM population with density greatly enhanced compared to the local halo density. On the other hand, this DM population has a reduced kinetic energy, making it difficult to detect in standard direct detection searches. The two methods we propose use energetic sources, in the form of underground nuclear accelerators or thermal sources, to upscatter the DM population accumulated in the Earth to make it easier to detect. We analyzed this scenario in a model-independent way, assuming only a point interaction between DM and nucleons, as well as in the context of a well-motivated model where DM interacts with the Standard Model through the exchange of a dark photon.

The first method we investigated for detecting strongly interacting DM is based on nuclear beams produced by underground accelerators, such as the LUNA~\cite{Prati:2020uxj} facility at LNGS. These accelerators feature beams of protons or light nuclei with kinetic energies in the $E_b \sim \mev$ range and currents near $I_b \sim 1\,\text{mA}$. When a beam nucleus strikes an ambient DM particle with mass $m_\chi \sim 1$--$100\,\gev$, the resulting DM energy typically falls into the sensitivity range of standard WIMP DM detectors. Based on realistic accelerator beam parameters and a modest detector proposal, we find potentially observable DM event rates for masses in the range $m_\chi \sim 1$--$100\,\gev$ for both a pointlike spin-independent interaction and for scattering mediated by a sub-GeV mass dark photon. This technique is able to probe parts of parameter space unconstrained by other searches, as seen in Figs.~\ref{fig:results} and~\ref{fig:DPplots}.

In the second method, ambient DM is upscattered by hot gas in a thermal source. This has the virtue of producing locally a very large flux of upscattered DM, but only with kinetic energies at the $\rm eV$ scale, much lower than the sensitivity range of standard WIMP detectors. However, proposed ultralow threshold detectors could soon be sensitive to DM in this energy range~\cite{Essig:2019kfe,Giomataris:2008ap,*Savvidis:2016wei,*NEWS-G:2017pxg,*Arnaud:2018bpc,SENSEI:2020dpa,SuperCDMS:2020ymb,CRESST:2019jnq,Hochberg:2015pha,Hochberg:2015fth,Schutz:2016tid,Hochberg:2017wce,Knapen:2017ekk}. Combined with the enormous effective fluxes of thermal sources and enhanced local DM densities in the Earth, this second approach could be a promising way to test strongly interacting DM in the near future.

\begin{figure}[htb]
    \centering
    \begin{tikzpicture}
    
    \draw[thick, -Triangle] (-0.2,0) -- node[font=\normalsize,below=18pt ]{energy (eV)} (6.9cm,0);

    \foreach \x in {0.7,1.4,2.1,2.8,3.5,4.2,4.9,5.6,6.3}
    \draw (\x cm,4pt) -- (\x cm,-4pt);

    \foreach \x/\descr in {2.1/1,4.2/10^3,6.3/10^6}
    \node[font=\normalsize, text height=1.75ex,
    text depth=.5ex] at (\x,-.5) {$\descr$};

    \draw[thick, -Triangle] (0,-0.2) -- node[font=\normalsize,left=35pt,rotate=90,xshift=20mm]{flux $\times$ volume (cm/s)
    } (0,5cm);

    \foreach \y in {1,...,4}
    \draw (4pt,\y cm) -- (-4pt, \y cm);

    \foreach \y/\descr in {1/10^{16},2/10^{22},3/10^{28},4/10^{34}}
    \node[font=\normalsize, text height=1.75ex,
    text depth=.5ex] at (-.6,\y) {$\descr$};
    
    \def\particles{(2,2) }
            \foreach \point in \particles{
            \foreach\i in {0,0.01,...,1} {
            \fill[opacity=\i*0.02,blue] \point ellipse ({0.5-\i} and {0.5-\i});
        }}
    \node[right,text=blue] at (2.5,2) {thermal sources};
        
    \def\particles{(6,1.1) }
            \foreach \point in \particles{
            \foreach\i in {0,0.01,...,1} {
            \fill[opacity=\i*0.02,green!60!black] \point ellipse ({1-2*\i} and {0.5-\i});
        }}
    \node[mynode,left,text=green!60!black, text width=2.2cm] at (5.3,1.1) {underground accelerators};
    
    \def\particles{(0.7,4) }
            \foreach \point in \particles{
            \foreach\i in {0,0.01,...,1} {
            \fill[opacity=\i*0.02,orange!80!black] \point ellipse ({0.5-\i} and {0.5-\i});
        }}
    \node[right,text=orange!80!black, text width=2.2cm] at (1.2,4) {underground detectors};

    \end{tikzpicture}
    \caption{Summary of beam energies and effective upscattering rates (beam flux $\times$ dark matter volume) for two acceleration schemes: underground nuclear accelerators (see Sec.~\ref{sec:acc}) and thermal sources (see Sec.~\ref{sec:hot}). For comparison, we also show a corresponding point for deep underground detectors encountering thermalized dark matter.}
    \label{fig:bigpic}
\end{figure}
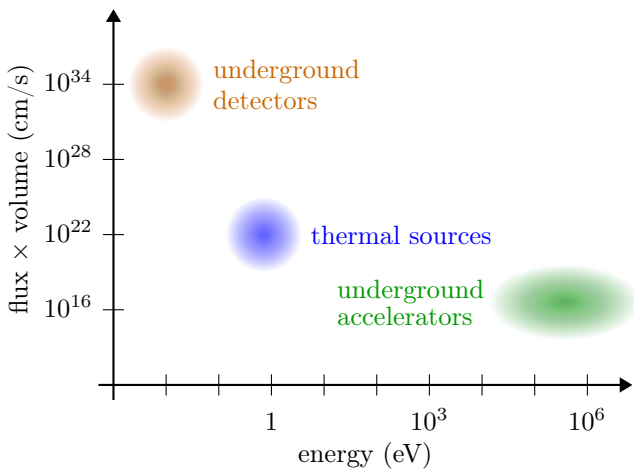

The two methods we have proposed can be characterized by: (i) the kinetic energy imparted to the upscattered DM; (ii) the effective rate of upscattering, which we define to be the flux of beam particles times the volume of DM impacted, corresponding to Eqs.~(\ref{eq:beamfluence}) and~(\ref{eq:LBfluence}). The ranges of these parameters is sketched in Fig.~\ref{fig:bigpic} for the two methods we have studied. These two methods of DM acceleration overlap well with two major thrusts in DM direct detection: searches for WIMP-like DM through nuclear scattering with energy deposition at the $\kev$ scale and searches for the scattering or absorption of light DM with energy transfers near the eV scale. Figure~\ref{fig:bigpic} offers a useful comparison of these techniques and also motivates studying other potential ways of accelerating and detecting strongly interacting DM for other combinations of energies and upscattering rates. For comparison, we also show in Fig.~\ref{fig:bigpic} a similar quantity for deep underground detectors surrounded by thermalized DM, where now the beam is identified with the detector volume moving through the surrounding DM at thermal velocity.

It is important to point out that the existing beam sources that we have considered are not optimized for maximizing the effective DM upscattering rate shown on the $y$ axis of Fig.~\ref{fig:bigpic}. It would be interesting to explore the feasibility of increasing this parameter at the expense of beam energy, collimation, and monochromaticity using other sources. A natural extension of beamlines and storage rings to higher densities are devices with plasma confinement with the main application being a nuclear fusion reactor. Temperatures up to 160 MK ($\approx$ 10 \textrm{keV}) have been reached and sustained for $\mathcal{O}\left(1000\right)$ seconds at the EAST reactor~\cite{song2014accomplishments} with ITER not too far behind~\cite{claessens2020iter}. The typical number densities required for fusion are in the range of $10^{20}~\textrm{m}^{-3}$ with plasma volume near $840~\textrm{m}^3$~\cite{Nature}. The viability of a low-threshold detector close to such a reactor and the impact of reactor shielding need to be analyzed in order to understand whether the large number of energetic particles confined in the plasma would help to offset much larger cosmogenic backgrounds existing on the Earth's surface to make it an efficient way of probing strongly interacting dark matter. One could also explore whether much more compact plasma retaining devices can be installed in underground laboratories with the main purpose of serving as a source of DM acceleration.  

Going to higher energy beams, ambient DM could be upscattered to higher kinetic energies, making it potentially easier to detect or distinguish from background. This could lead to signals in accelerator neutrino experiments with larger energy thresholds, albeit with larger volumes compared to ones considered in this work, in analogy to cosmic ray upscattered DM~\cite{Bringmann:2018cvk}. Examples of such setups include LSND~\cite{deNiverville:2011it,LSND:2001akn}, ISODAR~\cite{Alonso:2017fci}, and DUNE~\cite{DUNE:2015lol}. These setups could lead to sensitivity to larger DM masses owing to the superior ratio of beam energy to detection energy compared to accelerators considered in this work. The ISODAR proposal~\cite{Alonso:2017fci} looks especially promising in this respect, as it is planned to be installed at an underground laboratory with intrinsically low cosmogenic background.

\section*{Acknowledgements}

We thank Joseph Bramante, Ranny Budnik, Timon Emken, Andr\'ea Gaspert, Pietro Giampa, Gopolang Mohlabeng, Nirmal Raj, and Aaron Vincent for helpful discussions.
David McKeen and David Morrissey are supported by Discovery Grants from the Natural Sciences and Engineering Research Council of Canada~(NSERC). TRIUMF receives federal funding via a contribution agreement with the National Research Council~(NRC) of Canada. Marianne Moore acknowledges support by NSERC, the Fonds de Recherche du Qu\'ebec -- Nature et Technologies~(FRQNT) (Grant Nos. 273327 and 305494), and the Arthur Kerman Fellowship fund. Maxim Pospelov is supported in part by U.S. Department of Energy Grant No. desc0011842. Harikrishnan Ramani acknowledges the support from the Simons Investigator Award 824870, DOE Grant No. DE-SC0012012, NSF Grant No. PHY2014215, DOE HEP QuantISED Award No. 100495, and the Gordon and Betty Moore Foundation Grant No. GBMF7946.

\appendix

\section{Experimental Bounds on Strongly Interacting Dark Matter}
\label{sec:appb}

In the analysis above, we have presented exclusions on strongly interacting dark matter from direct detection experiments on the surface of the Earth and underground, measurements by satellite detectors, and cosmological observations. These bounds are typically presented in terms of SI DM with equal couplings to protons and neutrons that makes up the full DM abundance. In this appendix, we explain briefly how we generalize these bounds to a species $\chi$ that makes up only a fraction $f_\chi$ of the total DM density or that couples primarily to protons through a dark photon mediator. We also generalize previous analyses to include saturation of the nuclear cross section.

The experimental limits used in our study are the surface measurements by EDELWEISS~\cite{EDELWEISS:2019vjv} and CRESST~\cite{CRESST:2017ues}, the near-surface search by CDMS~\cite{CDMS:2002moo}, and deep underground searches by CDEX~\cite{CDEX:2019hzn,CDEX:2021cll}, CRESST-III~\cite{CRESST:2019jnq}, DarkSide-50~\cite{DarkSide:2018bpj}, and XENON1T~\cite{Aprile:2018dbl,XENON:2019gfn,XENON:2019zpr}. We also consider constraints derived from data from the XQC satellite~\cite{Wandelt:2000ad,Erickcek:2007jv,Mahdawi:2017cxz,Mahdawi:2018euy}, searches for cosmic ray upscattered DM~\cite{Cappiello:2019qsw,Bringmann:2018cvk} at PandaX-II~\cite{PandaX-II:2021kai}, and effects that strongly interacting DM would have on cosmological structure formation~\cite{Maamari:2020aqz,Buen-Abad:2021mvc,Rogers:2021byl}. Furthermore, we include bounds from the heating of helium in cryogenic Dewars~\cite{Neufeld:2018slx,Neufeld:2019xes,Xu:2021lmg} and the scattering-induced deexcitation of long-lived tantalum isomers~\cite{Pospelov:2019vuf,Lehnert:2019tuw} caused by DM captured in the Earth.

Direct DM searches typically present their (spin-independent) results as a lower exclusion limit on the per-nucleon cross section $\sigma_{\chi n}$. As the DM interaction with nuclei becomes stronger, DM will interact significantly with nuclei in the Earth, detector shielding, and atmosphere, causing it to slow down. Since realistic experiments have finite energy sensitivity, this leads to an upper exclusion limit on the interaction strength that they are sensitive to that depends on their energy threshold and detector location. A full treatment of this effect typically requires a detailed simulation of DM scattering in the experimental overburden~\cite{Mahdawi:2017cxz,Mahdawi:2017utm,Emken:2018run} that goes beyond the scope of this work. However, a simple estimate of overburden scattering based on a set of simplifying assumptions~\cite{Starkman:1990nj} (that are not always justified) has been found to give a reasonable approximation of the simulation results~\cite{Emken:2018run}, and we make use of it in our analysis. Specifically, in this estimate, DM particles are treated as traveling in a straight line from the halo to the detector, they are assumed to lose an average amount of energy in each scattering with nuclei along this path, and nuclear form factors are neglected. The condition for DM to be undetectable in a given experiment is then
\beq
E_{\chi}^{\rm th} \geq E_{\chi}^{i,\rm max}\,\exp(-d/\ell_{\rm eff}) \ ,
\label{eq:eover}
\eeq
where $E_{\chi}^{\rm th}$ is the lowest DM kinetic energy to which the detector is sensitive and depends on the detector energy threshold of the signal looked for (e.g. elastic nuclear scattering or the Migdal effect~\cite{Ibe:2017yqa,Dolan:2017xbu}), $E_{\chi}^{i,\rm max} = m_{\chi}(v_{\rm esc}+v_\text{eff})^2/2$ is the maximum DM kinetic energy in the halo, $d$ is the detector depth, and
\beq
\ell_{\rm eff}^{-1} = \frac{2}{d}\int\!dz\;\sum_N\,n_N(z)\,\frac{\mu_{\chi N}^2}{m_Nm_{\chi}}\,\sigma_{\chi N} \ ,
\eeq
where the integral runs over all nuclear species $N=(A,Z)$ in the overburden with local number density $n_N(z)$. Applying the condition of Eq.~\eqref{eq:eover} and using Eqs.~(\ref{eq:dsigder},\ref{eq:pernuc}) to relate nuclear cross sections to a per-nucleon value then leads to an upper exclusion of $\sigma_{\chi n}$. When possible, we take upper exclusion limits from published experimental results, but when they are not given, we estimate them using the condition of Eq.~\eqref{eq:eover}. 

We apply a further rescaling to generalize limits on DM that makes up the full cosmic abundance to a species that is only a fraction $f_\chi\leq 1$ of the total density. Given a lower exclusion of the per-nucleon cross section $\sigma_{n\chi}^{lo}$, reducing the DM fraction just decreases the expected number of DM events and the rescaling is simply
\beq
\sigma_{n\chi}^{lo} \to \sigma_{n\chi}^{lo}/f_\chi \ .
\eeq
For upper exclusions of the per-nucleon cross $\sigma_{\chi n}^{hi}$, the dependence of the event rate on the DM-nucleon cross section is much steeper than linear~\cite{Emken:2018run}. Indeed, within the threshold approximation of Eq.~\eqref{eq:eover}, the local DM density does not enter at all. As a result, we do not apply a rescaling by $f_\chi$ to upper exclusions $\sigma_{n\chi}^{hi}$. The collected exclusions derived using these methods for SI DM are summarized in Fig.~\ref{fig:SIDMbounds} for $f_\chi = 1,\,10^{-3},\,10^{-6},\,10^{-9}$.

\begin{figure*}[ttt]
\centering
\begin{minipage}[b]{0.5\textwidth}
        \centering
        \includegraphics[height=5.705cm]{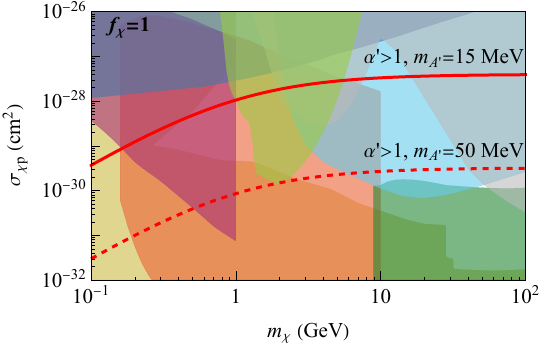} 
  \end{minipage}
  \begin{minipage}[b]{0.5\textwidth}
    \centering
    \includegraphics[height=5.705cm]{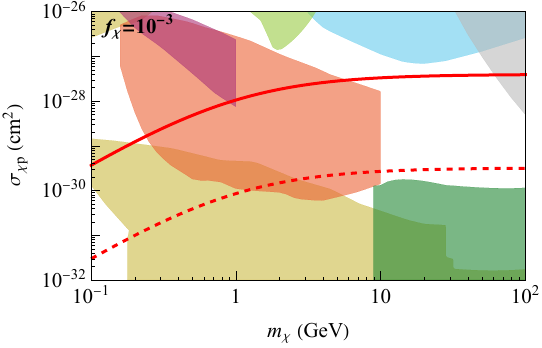}
  \end{minipage} 
\begin{minipage}[b]{0.5\textwidth}
        \centering
        \includegraphics[height=5.705cm]{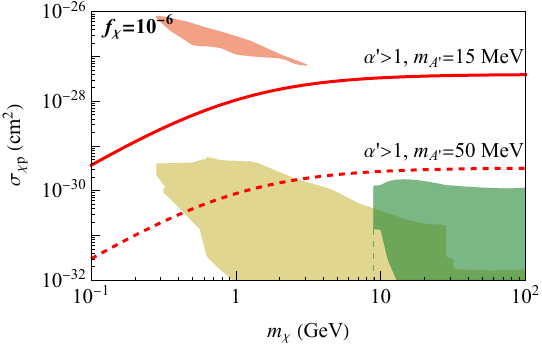} 
  \end{minipage}
  \begin{minipage}[b]{0.5\textwidth}
    \centering
    \includegraphics[height=5.705cm]{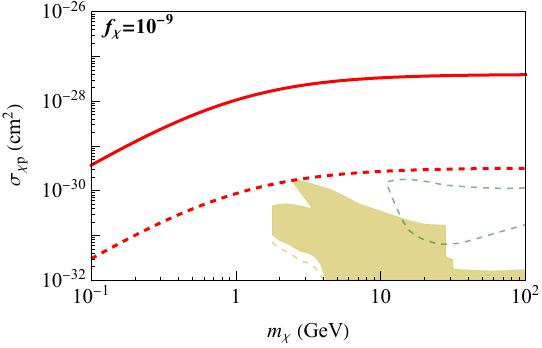}
  \end{minipage} 
  \includegraphics[width=.73\textwidth]{figs/legend.pdf}
\caption{Current bounds on dark photon mediated dark matter that makes up a fraction $f_\chi = 1,10^{-3}, 10^{-6}, 10^{-9}$ of the total DM density as a function of its mass and effective per-nucleon cross section for dark photon masses of $m_{A'}=15,\,50\,\mev$. The solid filled regions are excluded for both dark photon masses, while the dashed regions are only ruled out for $m_{A'}=50\,\mev$. The dashed~(solid) red lines show the regions above which $\alpha' > 1$ is needed to obtain the corresponding DM-proton cross for $m_{A'} = 15\,\mev$~($50\,\mev$). Details of how these bounds were obtained and a list of the original experimental sources are given in Appendix~\ref{sec:appb}.
}
\label{fig:DPexclusions}
\end{figure*}

The prescription we use here can also be generalized to DM for which the connections between the nuclear and nucleon cross sections of DM do not follow exactly Eqs.~(\ref{eq:dsigder},\ref{eq:pernuc}). Given a bound on the per-nucleon cross section for minimal SI DM, it can be converted into a limit on the relevant nuclear cross sections and then recast as a limit on an effective per nucleon cross section for other scenarios. We use this approach to add the effect saturation of the nuclear cross section as described in Eq.~\eqref{eq:sigeff}, which turns out to be relevant only for some upper bounds from surface experiments and XQC. We also apply this method to derive limits on dark photon-mediated DM that couples almost exclusively to protons. The limit on $\sigma_{\chi p}$ for such DP DM given a bound on $\sigma_{\chi n}$ for minimal elastic nuclear scattering of SI DM is then
\beq
\begin{aligned}
\sigma_{\chi p}^{\rm lim} &= \sigma_{\chi n}^{\rm lim}\lrf{A}{Z}^2
\\
&\times \left(
\int\!dq^2\;\eta\,|F_N|^2\bigg{/}
\int\!dq^2\;\eta\,|F_N|^2|F_{\chi}|^2 \right) \ ,
\end{aligned}
\eeq
where $\eta = \eta(v_{\rm min})$ is the usual halo function~\cite{Bertone:2004pz}, $F_N$ is the nuclear SI form factor~\cite{Engel:1992bf,Lewin:1995rx}, and $F_{\chi}$ is the DM mediator form factor given in Eq.~\eqref{eq:ffdp}. The exclusions on dark photon mediated DM derived with these methods are shown in Fig.~\ref{fig:DPexclusions} for DM fractions of $f_{\chi} = 1,\,10^{-3},\,10^{-6},\,10^{-9}$ and dark photon masses $m_{A'}=15,\,50\,\mev$. The shaded regions in both panels are excluded for $m_{A'} = 15\,\mev$, while the regions inside the dashed contours are excluded only for $m_{A'}=50\,\mev$.

\section{Calculating the Distribution of Thermally Upscattered Dark Matter}
\label{sec:appa}
This appendix contains our calculations of the DM energy distribution produced by scattering with hot SM particles in a thermal source that we use in Sec.~\ref{sec:hot}. While we show the approximate distribution in Eq.~(\ref{eq:dm-lbdist}), we use the more exact treatment below to generate the event rates that go into Fig.~\ref{fig:bulbrates}.

As a starting point, we consider strongly interacting DM that has thermalized with the local environment on Earth. We take it to be nonrelativistic with a Maxwell-Boltzmann phase space distribution,
\begin{equation}
\begin{aligned}
f_\chi(k)=\left(\frac{2\pi}{m_\chi T_\chi}\right)^{3/2}\exp\left(-\frac{k^2}{2m_\chi T_\chi}\right)
\end{aligned}
\label{eq:fchiMB}
\end{equation}
and a temperature $T_\chi\sim 300~\rm K$ determined by the local thermodynamic temperature. Some of this DM encounters SM gas particles near a thermal source at a high temperature $T_g\gg T_\chi$. For our purposes we assume that these gas particles achieve thermal equilibrium with the thermal source. While they are comparatively hot, these particles' phase space distribution, $f_g$, can also be approximated by a Maxwell-Boltzmann distribution as in~(\ref{eq:fchiMB}) with $m_\chi\to m_g$, $T_\chi\to T_g$. Given a matrix element for DM to elastically scatter on the gas of $\cal M$, the rate of DM scattering is
\begin{equation}
\begin{aligned}
\Gamma&=n_g\int\frac{d^3\boldsymbol k}{2m_\chi(2\pi)^3}f_\chi(k)\frac{d^3\boldsymbol p}{2m_g(2\pi)^3}f_g(p)\frac{d^3{\boldsymbol k}^\prime}{2m_\chi(2\pi)^3}
\\
&\quad\quad\times\frac{d^3{\boldsymbol p}^\prime}{2m_g(2\pi)^3}(2\pi)^4\delta^{(4)}({\boldsymbol k}+{\boldsymbol p}-{\boldsymbol k}^\prime-{\boldsymbol p}^\prime)\left|{\cal M}\right|^2
\end{aligned}
\end{equation}
where $n_g$ is the number density of gas particles. $\boldsymbol k$ and $\boldsymbol p$ label the DM and gas momenta and primes denote final state quantities.

Using the delta functions to integrate over the unobserved final state gas momentum and reexpressing the initial momenta in terms of the center-of-mass velocity $\boldsymbol V$ and relative velocity $\boldsymbol v$,
\begin{equation}
{\boldsymbol k}=m_\chi {\boldsymbol V}+\mu_{\chi g} {\boldsymbol v},~{\boldsymbol p}=m_g {\boldsymbol V}-\mu_{\chi g} {\boldsymbol v},
\end{equation}
allows us to write the rate as
\begin{widetext}
\begin{equation}
\begin{aligned}
&\Gamma=\frac{n_g \mu_{\chi g} m_g}{8(2\pi)^5}\int\,V\,dV\,v^2\,dv\,d\cos\theta \,{k^\prime}\,dk^\prime\,d\cos\theta^\prime \times f_\chi(k)f_{\rm gas}(p)\delta\left(\cos\theta^\prime-\frac{{k^\prime}^2+m_\chi^2V^2-\mu_{\chi g}^2v^2}{2m_\chi V k^\prime}\right)\left|{\cal M}\right|^2
\end{aligned}
\end{equation}
\end{widetext}
where $\theta$ ($\theta^\prime$) is the angle between $\boldsymbol V$ and $\boldsymbol v$ (${\boldsymbol k}^\prime$). If we make the simplifying assumption that the matrix element is independent of the momenta then we can perform these integrations to find the total rate:
\begin{equation}
\Gamma=\frac{n_g\left|{\cal M}\right|^2}{4\sqrt{2\pi^3}\left(m_\chi +m_g\right)^2}\sqrt{\frac{T_g}{m_g}+\frac{T_\chi}{m_\chi}}.
\end{equation}

The normalized flux distribution of the upscattered DM flux with respect to its kinetic energy, $E_\chi={k^\prime}^2/(2m_\chi)$, is then
\begin{align}
\frac{1}{\Phi_\chi}\frac{d\Phi_\chi}{dE_\chi}&=\frac{1}{\Gamma}\frac{m_\chi}{k^\prime}\frac{d\Gamma}{dk^\prime}=\frac{m_g m_\chi}{4\mu_{\chi g}^2\left(T_g-T_\chi\right)}
\nonumber\\
&\times\Bigg\{\frac{b_+}{\sqrt{b_+^2+1}}\exp\left(-\frac{a}{b_+^2+1}\right){\rm erf}\left(\sqrt{\frac{a}{b_+^2+1}}b_+\right)
\nonumber\\
&-\frac{b_-}{\sqrt{b_-^2+1}}\exp\left(-\frac{a}{b_-^2+1}\right){\rm erf}\left(\sqrt{\frac{a}{b_-^2+1}}b_-\right)\Bigg\}
\end{align}
with
\begin{equation}
\begin{aligned}
a&\equiv\frac{E_\chi}{m_\chi}\left(\frac{m_\chi}{T_\chi}+\frac{m_g}{T_g}\right),~b_-\equiv\sqrt{\frac{m_g T_\chi}{m_\chi T_g}},
\\
b_+&\equiv b_-\left(1+\frac{2m_\chi}{T_\chi}\frac{T_g-T_\chi}{m_\chi +m_{\rm gas}}\right).
\end{aligned}
\end{equation}
In the limit $T_\chi\to 0$ this reduces to the expression in Eq.~(\ref{eq:dm-lbdist}).

\vspace{0.8cm}

\bibliography{biblio}

\end{document}